\documentclass[sigconf]{acmart}

\AtBeginDocument{%
  \providecommand\BibTeX{{%
    \normalfont B\kern-0.5em{\scshape i\kern-0.25em b}\kern-0.8em\TeX}}}

\copyrightyear{2026}
\acmYear{2026}
\setcopyright{cc}
\setcctype{by}
\acmConference[CHI '26]{Proceedings of the 2026 CHI Conference on Human Factors in Computing Systems}{April 13--17, 2026}{Barcelona, Spain}
\acmBooktitle{Proceedings of the 2026 CHI Conference on Human Factors in Computing Systems (CHI '26), April 13--17, 2026, Barcelona, Spain}
\acmPrice{}
\acmDOI{10.1145/3772318.3791865}
\acmISBN{979-8-4007-2278-3/2026/04}

\usepackage{xcolor}
\usepackage{multirow}
\usepackage{xspace}
\usepackage{booktabs}
\usepackage{subfigure}

\newcommand{\ie}{\emph{i.e.,}\xspace}
\newcommand{\eg}{\emph{e.g.,}\xspace}

\newtheorem{finding} {Finding}

\begin{document}

\title{Hesitation and Tolerance in Recommender Systems}

\author{Kuan Zou}
\authornote{Kuan Zou was also affiliated with Alibaba Cloud Computing when this work was conducted.}
\affiliation{
  \institution{Nanyang Technological University}
  \city{Singapore}
  \country{Singapore}
}
\email{zouk0002@e.ntu.edu.sg}

\author{Aixin Sun}
\authornote{Corresponding author.}
\affiliation{
  \institution{Nanyang Technological University}
  \city{Singapore}
  \country{Singapore}
}
\email{axsun@ntu.edu.sg}

\author{Yitong Ji}
\affiliation{
 \institution{Nanyang Technological University}
 \city{Singapore}
 \country{Singapore}
}
\email{s190004@e.ntu.edu.sg}

\author{Hao Zhang}
\affiliation{
 \institution{Nanyang Technological University}
 \city{Singapore}
 \country{Singapore}
}
\email{hao007@e.ntu.edu.sg}

\author{Jing Wang}
\affiliation{
  \institution{Nanyang Technological University}
  \city{Singapore}
  \country{Singapore}
}
\email{jing005@e.ntu.edu.sg}

\author{Zhuohao (Jerry) Zhang}
\affiliation{
  \institution{University of Washington}
  \city{Seattle}
  \country{United States}
}
\email{zhuohao@uw.edu}

\author{Xuemeng Jiang}
\affiliation{
  \institution{Alibaba Digital Media \& Entertainment Group}
  \city{Beijing}
  \country{China}
}
\email{xuemeng.jxm@alibaba-inc.com}

\renewcommand{\shortauthors}{Zou and Sun, et al.}

\begin{abstract}
Users' interactions with recommender systems often involve more than simple acceptance or rejection. We highlight two overlooked states: \textit{hesitation}, when people deliberate without certainty, and \textit{tolerance}, when this hesitation escalates into unwanted engagement before ending in disinterest. Across two large-scale surveys ($N=6,644$ and $N=3,864$), hesitation was nearly universal, and tolerance emerged as a recurring source of wasted time, frustration, and diminished trust. Analyses of e-commerce and short-video platforms confirm that tolerance behaviors, such as clicking without purchase or shallow viewing, correlate with decreased activity. Finally, an online field study at scale shows that even lightweight strategies treating tolerance as distinct from interest can improve retention while reducing wasted effort. By surfacing hesitation and tolerance as consequential states, this work reframes how recommender systems should interpret feedback, moving beyond clicks and dwell time toward designs that respect user value, reduce hidden costs, and sustain engagement.
\end{abstract}



\maketitle

\section{Introduction}
\label{sec:intro}

Imagine opening a recommended video that seems appealing at first. You pause, watch for a while, and then realize it is not what you wanted. The time and attention you spent are gone, an experience of disappointment rather than value. Such moments are common, yet most recommender systems (RecSys) log them simply as a ``click,'' treating them as success.

Over the past decade, platforms from e-commerce~\cite{10.1145/3292500.3330652,10.1145/1297231.1297255} to short-video streaming~\cite{zannettou2024analyzing,10.1145/3580305.3599922,10.1145/3627508.3638341}  have invested heavily in recommender systems  to increase retention and engagement~\cite{du2019improve,10.1145/3341215.3356261,10.1145/3543873.3584628,10.1145/3306309.3306322,jannach2019measuring}. While these systems are celebrated for their reach and profitability, the signals that drive them remain strikingly coarse. For example, clicks are routinely collapsed into binary categories of \textit{positive} (interested) or \textit{negative} (disinterested) in recommender systems, erasing the lived nuances of curiosity, doubt, and regret before ``click''. As a result, metrics like Click-Through Rate (CTR) have become the dominant proxy for user satisfaction~\cite{10.1145/3539618.3591755,zhang2014explicit,10.1145/2783258.2783273}, even though they frequently capture hesitation or frustration rather than genuine preference. 

The complexity and diversity of user intentions, and even emotions, complicate the correlation between intentions and the limited set of behavior types~\cite{xia2021graph}, which is deeply problematic. Users do not simply react to algorithms; they strategically explore, second-guess, and occasionally persist with content they do not truly want. These moments are costly (not in computation, but in human terms): time wasted, trust eroded, agency undermined. Yet they remain invisible in current optimization frameworks.

Addressing this gap requires shifting attention from model architectures to the dynamics of \textit{user-recommender interaction}. 
Decades of HCI research demonstrate that users act strategically, form \textit{folk theories} of algorithms, and actively probe or steer them~\cite{eslami2016first,rader2015understanding,devito2017algorithms}. 
Perspectives on agency, trust, and cognitive load further caution that engagement metrics may signal doubt, frustration, or compulsion rather than satisfaction~\cite{lukoff2021design,pirolli2007cognitive}. 
These insights suggest that current RecSys models overlook critical \textit{interaction states} between interest and disinterest, such as moments of hesitation, second-guessing, and persistence despite dissatisfaction.

In this paper, we investigate two such states. 
\textit{Hesitation} arises when users deliberate over recommended content, clicking without certainty.  
\textit{Tolerance} occurs when hesitation escalates into unwanted engagement, leading to frustration and wasted effort. 
These states are consequential because they introduce hidden \textit{interaction costs}, time, cognitive effort, and negative emotions, that erode trust and retention while remaining invisible to current optimization metrics. By examining these states, we aim to broaden how recommender systems interpret feedback, shifting attention from maximizing clicks to minimizing wasted effort.
To investigate these dynamics, we pose three research questions: \textbf{RQ1:} Does an intermediate state exist between interest and disinterest in recommender interactions?  \textbf{RQ2:} If so, how can this state be systematically identified and modeled?   \textbf{RQ3:} If captured, can recommender systems leverage these signals to improve user retention?  

We conducted a multi-method, multi-scale study comprising 
(1) large-scale surveys, 
(2) behavioral log analyses, and 
(3) a field deployment. 
For \textbf{RQ1}, we designed two surveys ($N=6,644$ and $N=3,864$). 
Findings show that: (i) users indeed experience an ambiguous state, which we term \textit{hesitation}; 
(ii) hesitation, when coupled with unproductive costs and eventual disinterest, often develops into \textit{tolerance}; 
and (iii) tolerance undermines enthusiasm, reduces activity, and may ultimately lead to abandonment. 
From these survey findings, we characterized tolerance signals and demonstrated their potential for recommender optimization, partially addressing \textbf{RQ2}.  

To further address \textbf{RQ2} and \textbf{RQ3}, we analyzed behavioral datasets from e-commerce and short-video platforms. 
Results indicate that higher tolerance correlates with reduced activity. 
Finally, we validated our approach through large-scale online A/B testing on a video-sharing platform with millions of users. The results demonstrate that even lightweight strategies, such as relabeling tolerance signals, can produce measurable improvements in next-day retention. Training models with these signals not only boosts activity but also enhances retention, offering a cost-effective response to long-standing industrial challenges.

Across multiple deployments, these adjustments yielded consistent gains, with next-day retention increasing by up to \textbf{0.67\%}, a figure that, while modest, represents a substantial improvement at industrial scale. Beyond technical optimization, our work reframes hesitation and tolerance as consequential interaction states. By doing so, we move recommender system research toward a more human-centered perspective, highlighting how systems can respect users' time, preserve trust, and foster sustainable engagement.

Our contributions are fourfold: 
(1) we conceptualize and operationalize \textit{hesitation} and \textit{tolerance} as measurable intermediate interaction states, empirically establishing their prevalence and \textit{direct impact on retention}; 
(2) we provide converging evidence, across surveys, behavioral logs, and an industrial-scale A/B test, that these states depress trust, satisfaction, and engagement; 
(3) we propose and validate modeling strategies that incorporate tolerance signals as weak positives/negatives, bridging RecSys optimization with user-centered concerns about user cost and agency; and 
(4) we demonstrate practical impact: modest relabeling of tolerance-related signals both improves retention and reduces wasted effort, offering a pathway to design recommenders that respect user time and foster sustainable engagement.
\section{Related Work}

Research in user behavior modeling and optimization objectives in recommender systems exist in both RecSys and SIGCHI communities. User behaviors, such as clicks, views, and purchases, serve as essential data for the recommender system to predict user preferences and satisfaction. Leveraging user behaviors, various optimization objectives have been proposed, \ie predictions of click-through rate, click conversion rate, dwell time, and even long-term outcomes such as user retention. In the meantime, prior work have examined how users understand, experience, and adapt to recommender systems, focusing less on algorithmic optimization and more on users' mental models, agency, and the social and psychological consequences of interacting with algorithmic recommendations. In this section, we discuss existing work related to user behavior modeling in both RecSys and SIGCHI research, as well as optimization objectives in recommender systems (RecSys)
\subsection{Modeling User Behaviors in RecSys}
Early recommender systems research often emphasized how users form and act on their \textit{mental models} of algorithms. Studies have shown that people rarely treat recommendations as neutral suggestions; instead, they construct ``folk theories'' of how algorithms operate and attempt to steer or correct them through their actions. For example, users may deliberately click, like, or hide content to signal preferences, even when these actions do not align with their actual interest, reflecting a belief that such signals will shape future recommendations~\cite{kulesza2012tell, eslami2016first, rader2015understanding, devito2017algorithms}. This line of work highlights that user behaviors in recommender systems cannot be interpreted as direct expressions of preference, but are deeply entangled with users' understandings, expectations, and strategies for influencing algorithms. Beyond conceptual models, prior work has also examined hesitation as a micro-level social signal manifested through nonverbal cues such as hand movements and gaze dynamics in conversational recommendation settings~\cite{10.1007/s11042-014-1933-2}.

Building on this perspective, subsequent user-centered studies have examined how people decide whether to accept or reject recommendations, especially in contexts like news feeds, video streaming, or e-commerce. Research shows that users often deliberate before acting on a recommendation, weighing factors such as trust in the platform, perceived relevance, and anticipated consequences of engagement~\cite{wang2022will, lukoff2021design}. When users do click, these behaviors are not always equivalent to satisfaction or endorsement; instead, actions like briefly opening a video, skipping ahead, or abandoning a purchase may signal ambivalence or doubt~\cite{chaudhary2022unintended, zannettou2024analyzing}. Work in this area has drawn on broader user-centered theories of agency, cognitive load, and information foraging to explain why clicks and views can mask underlying uncertainty or disengagement~\cite{pirolli2007cognitive, card2001information}.

Complementing these user-centered perspectives, recommender systems research has long treated user behavior as the primary input for training and evaluating models. User clicks, views, and purchases serve as essential signals for predicting preferences and satisfaction, and a large body of work has refined optimization objectives around metrics such as click-through rate, conversion rate, dwell time, and retention \cite{wu2022graph,chen2023bias,wang2021survey,wang2023learning,frolov2016fifty,park2022exploiting,zhao2018recommendations,liu2010personalized}. While these behavioral proxies have proven useful, they often treat user actions as direct indicators of interest, overlooking the more nuanced and long-tail signals that may reflect ambivalence, disengagement, or emotional states \cite{marko2018emotions,xia2021graph}. For instance, preliminary work has explored the impact of emotions and personality on recommender systems \cite{marko2018emotions}, while more recent efforts have examined how user interests \cite{zannettou2024analyzing}, activities \cite{pan2023learning,li2022modeling}, and temporal signals such as dwell time \cite{zhan2022deconfounding,zhao2024counteracting} shape system performance and evaluation. 

Another stream of research has focused on mitigating systematic and algorithmic biases in user behavior modeling. Systematic biases are often addressed by promoting equity of attention for items or improving model performance across subgroups \cite{beutel2019fairness,biega2018equity,mehrotra2018towards,zehlike2020reducing}, while algorithmic biases are increasingly approached from the perspective of causal reasoning \cite{pearl2012the}. This has led to the development of causal embedding \cite{bonner2018causal,guo2019pal,zhang2023disentangled}, inverse propensity weighting \cite{christakopoulou2020deconfounding,kool2020ancestral,bietti2021a}, and causal intervention \cite{wang2021deconfounded,yang2021topn,zhang2021causal} methods to correct confounding factors and improve robustness. Beyond bias mitigation, researchers have also explored mining richer user features \cite{cho2019no,park2017rectime,xiao2022training} and designing multi-task frameworks \cite{chang2023pepnet,ma2018modeling,tang2020progressive,zhao2019recommending} to improve personalization and enhance user satisfaction.

Despite these advances, both HCI and RecSys research tend to focus on either users' conceptual models of algorithms or measurable optimization objectives tied to observable behaviors. Less attention has been given to the dynamic decision-making processes that precede or underlie these behaviors. In particular, prior work has not systematically addressed the dynamics of \textit{hesitation}, when users deliberate at the threshold of action, or \textit{tolerance}, when they continue engaging with content they do not truly want. Our research introduces and formalizes these two critical concepts, filling a gap in both academic and industry literature by offering foundational insights into subtle user decision-making patterns that existing models and theories have largely overlooked.

\subsection{Optimization Objectives in RecSys}

Optimizing recommender systems is a complex endeavor that necessitates considering multiple facets and perspectives, and it is accompanied by trade-offs between evaluation metrics. A comprehensive evaluation must address both system-centric aspects and user-centric aspects. Furthermore, a strong focus, particularly in industry, is placed on business effectiveness and overall goals, such as increasing revenue and customer engagement~\cite{zangerle2022evaluating,zou2025survey}.

Click is one of the key user behaviors in recommender systems. The conversion of a click can lead to an item purchase or an advertisement view on an online platform, ultimately generating business revenue. Accurately predicting whether a user will click on an item in a given context is crucial for optimizing how items are displayed to maximize business outcomes~\cite{ZhouZSFZMYJLG18DIN, ZhouMFPBZZG19DIEN, LinQGDT0023map}. This task, known as click-through rate (CTR) prediction, has been extensively studied, with numerous models proposed over the years. 

Conventional CTR models, such as Factorization Machines (FM; \cite{Rendle10FM}), model pairwise feature interactions for binary classification of clicks and non-clicks. Other approaches~\cite{GuoTYLH17deepFM, LianZZCXS18xDeepFM} extend these models by incorporating higher-order feature interactions through the use of deep neural networks (DNNs).~\citet{ZhouZSFZMYJLG18DIN} further enhance CTR models by introducing a local activation unit to capture the relevance between candidate ads and users' historical behaviors. In DIEN~\cite{ZhouMFPBZZG19DIEN}, the authors propose leveraging auxiliary loss in CTR models to supervise the learning of user behavior sequences at each time step. 

However, traditional CTR prediction methods typically treat all clicks equally, without distinguishing between the \textit{quality} of clicks. Studies have shown that while clicks reflect users' initial impressions, they do not necessarily indicate user satisfaction with items after further exploration~\cite{XieMZ0L23reweightingClicksDwellTime, WenYE19postClick}. In HCI research, this distinction aligns with findings that actions such as clicking, skipping, or briefly viewing may stem from ambivalence or testing the system rather than genuine interest~\cite{lukoff2021design,chaudhary2022unintended}. Together, these insights suggest that post-click behaviors should be considered in CTR models to better capture the nuance of user preferences. 

Beyond CTR, click conversion rate (CVR) prediction is also a crucial task in recommender systems. Generally, click conversion relates to the ``click-to-purchase'' path, which directly impacts final revenue. However, predicting CVR is fundamentally more challenging than predicting CTR due to issues such as selection bias and the sparsity of positive samples in the data. To address these challenges,~\citet{MaZHWHZG18entireSpace} propose a multi-task learning approach that simultaneously trains CTR and CVR models, allowing the CVR prediction task to benefit from the abundant positive samples available in the CTR prediction task. Similarly, the authors of~\cite{WenZWLBLY20PostClickDecomposition} introduce a multi-task model based on conditional probability rules to enhance CVR learning. \citet{WenZLBWC21CVR} further propose to decompose user behaviors into detailed actions, such as clicking on item images and engaging in conversations with sellers, providing richer signals for CVR prediction. Such decompositions echo work on understanding the micro-decisions that shape users' mental models of systems and influence their willingness to trust or act on recommendations \cite{rader2015understanding,eslami2016first}.

When it comes to video recommendation scenarios, beyond clicks, watch-time prediction is regarded as a more representative metric of user engagement~\cite{wu2018beyond,de2021extracting,covington2016deep}. \citet{covington2016deep} presents an industry-standard solution to watch-time prediction, where the regression problem is transformed into the weighted logistic regression. \citet{zhan2022deconfounding} further analyze the duration bias in watch-time prediction via a causal graph and a duration-deconfounded quantile-based framework. \citet{zhao2024counteracting} introduce the counterfactual watch time concept to measure the potential duration bias and devise a corresponding counterfactual watch model to solve this problem. Prior work has questioned whether watch time or dwell time truly equate to satisfaction, pointing to cases where prolonged engagement may actually reflect frustration, compulsion, or lack of alternatives \cite{lukoff2021design,pirolli2007cognitive}. 

In addition to click-based objectives, a wide range of interaction signals are used in both academic and industrial recommender systems. Long-CTR, which counts only clicks that lead to sustained consumption, has long been deployed in video and content platforms as a more robust engagement signal than raw clicks~\cite{10.1145/1864708.1864770,jadon2024comprehensive}. Beyond this, completion-based measures—such as watched-percentage or scrolled-percentage ratios—are widely adopted using distributional labeling or causal debiasing techniques~\cite{zhan2022deconfounding,zhang2023leveraging}. Low-engagement indicators, including skip rate and early exits, are frequently leveraged to capture disengagement in short-video recommenders~\cite{gong2022real}.
At the sequence level, session-duration dynamics and next-item transitions are commonly modeled through long-sequence or session-based architectures~\cite{si2024twin,wu2024learned,mei2022lightweight}. Platforms also rely on explicit feedback such as likes, saves, and ratings to refine user preferences~\cite{qin2021bootstrapping,shao2024optimizing}. Finally, novelty and serendipity-oriented objectives are used to encourage exploratory consumption and mitigate fatigue~\cite{li2020purs,lin2022feature,wang2023industrial}.
These metrics collectively illustrate the breadth of signals used to assess engagement and value in practice, and they highlight the limitations of interpreting surface-level behaviors as direct indicators of interest—motivating the need to examine interaction-level states such as hesitation and tolerance.

Our work fundamentally differs from the aforementioned papers. Rather than focusing on immediate feedback like clicks and dwell time, we examine users' decision-making processes, particularly their emotional responses to items after clicking. Building on this, we explore a long-term business metric: user retention. In the following section, we review existing research on optimizing user retention in recommender systems.

\subsection{User Retention Optimization}

User retention is typically defined as the ratio of users who revisit a system after their initial interaction~\cite{cai2023reinforcing}. It has become an increasingly critical metric in the evaluation and optimization of recommender systems~\cite{wu2017returning,zou2019reinforcement,cai2023reinforcing,xue2023resact,silveira2017how,zhang2021user,10.1145/3604915.3608818,gomez2016netflix}. Unlike immediate feedback, such as point-wise~\cite{10.1145/3447548.3467205,10.1145/3336191.3371818} and list-wise~\cite{10.1145/3404835.3463115,10.1145/3397271.3401330} signals, which reflect user responses after a single interaction, retention captures long-term feedback resulting from multiple interactions over time~\cite{cai2023reinforcing}. This longer horizon aligns with prior findings that emphasize sustained user experiences and trust rather than single clicks or sessions.

Due to the nature of delayed feedback, reinforcement learning (RL) methods are particularly well-suited to optimize user retention by addressing long-term rewards and user engagement~\cite{10.1145/3534678.3539040,10.1145/3580305.3599473,zou2019reinforcement}. One of the representative works is RLUR~\cite{cai2023reinforcing}, designed to address challenges such as uncertainty, bias, and delayed feedback, thereby improving retention by focusing on long-term positive outcomes. In addition, some studies have explored optimizing retention through RL by modeling mid- and long-term user behaviors~\cite{wu2017returning,zou2019reinforcement,xue2023resact}, investigating interpretable aspects of retention~\cite{10.1145/3604915.3608818}, or integrating exploratory behaviors with reinforcement learning to enhance user experiences~\cite{10.1145/3460231.3474236}. Recent efforts~\cite{kohavi2024false} also analyze the impact of false positives in online A/B tests for recommender systems, where the mined false positives can be regarded as retention signals.

While technical approaches frame retention as a long-term optimization objective, prior work has shown that returning to a system is not always a proxy for satisfaction: users may return out of habit, compulsion, or lack of alternatives rather than genuine engagement \cite{lukoff2021design,devito2017algorithms}. Thus, combining retention-oriented optimization with research on user agency, motivation, and trust highlights the importance of distinguishing between surface-level engagement and meaningful, value-aligned use. We focus on this synthesis by introducing hesitation and tolerance as behavioral signals that help bridge immediate and long-term perspectives on user-recommender interactions.

\section{User Interaction States in RecSys}
\label{sec:userStudy}

Hesitation and tolerance are inherently subjective states, involving psychological costs, emotional responses, and perceptions of wasted effort. In this paper, we define \textbf{hesitation} as \textit{a user state involving uncertainty and additional cognitive effort before deciding interest}, and \textbf{tolerance} as \textit{a user state in which initial curiosity leads to unwanted engagement that users continue despite realizing disinterest}. Such experiences are difficult to capture in behavioral logs, which record only observable actions. 

Therefore, to address our first research question: whether an intermediate state exists between interest and disinterest in recommender systems, we conducted two large-scale online surveys. Surveys are well-suited to capture such psychological states, which are otherwise invisible in behavioral traces, and to establish their prevalence and subjective costs. Our goal was to examine everyday user experiences with recommender systems, focusing on whether intermediate interaction states can be identified, how users describe them, and what consequences they report.
Rather than presenting participants with predefined labels such as hesitation or tolerance, we employed scenario-based questions to elicit their experiences in concrete forms (\eg clicking without purchasing, speed-watching a video). The first survey established behavioral signals and validated our question framing, while the second replicated these measures with an independent sample and added items probing longer-term consequences such as frustration, reduced activity, and attrition. Together, the two surveys provide complementary evidence for understanding intermediate user-recommender interaction states.

We use \textbf{tolerance} in its dictionary sense of ``enduring or putting up with something undesirable,'' rather than \textit{the prosocial or interpersonal meaning of tolerance}. We adopt this term because users frequently described experiences in which they continued engaging with content despite realizing it was not what they wanted, often accompanied by feelings of wasted time or frustration. Other similar alternatives include ``unwanted engagement'' and ``regretful engagement,'' and we chose ``tolerance'' to best capture the pattern of brief, low-value persistence observed in our data.

\subsection{Method}
We introduce how we conducted the two large-scale surveys. Both surveys were reviewed and approved by Alibaba's internal research ethics and compliance process.

\begin{figure*}
    \centering
    \subfigure[Respondents of the first survey]
    {
        \label{fig:1st_round_demo}    
        \includegraphics[trim={0cm 0cm 0cm 0cm},clip,width=0.75\textwidth]{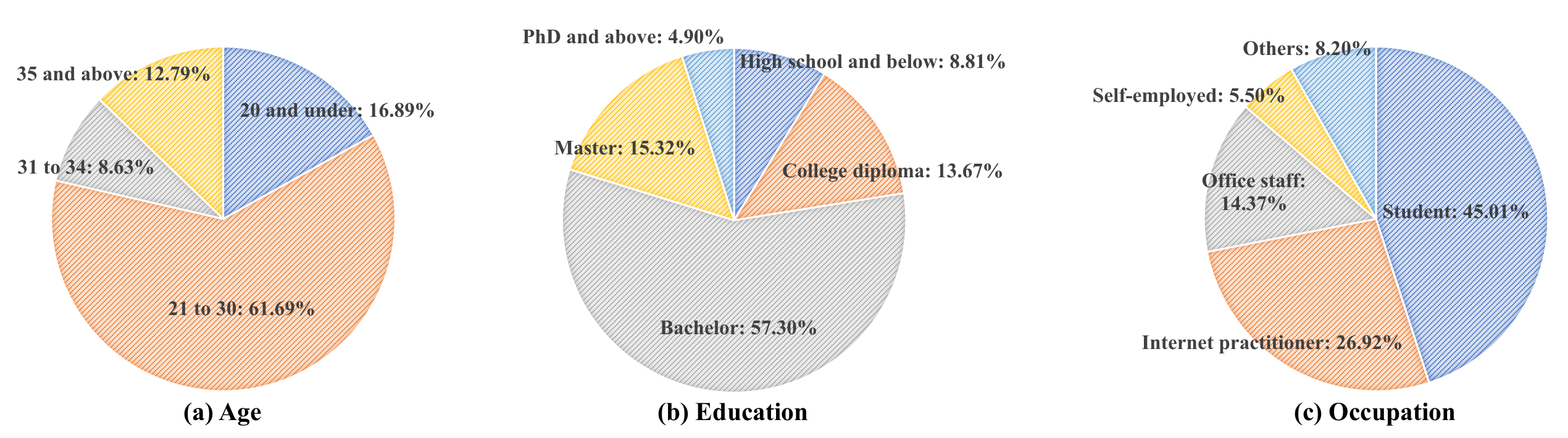}
    }
    \subfigure[Respondents of the second survey]
    {
        \label{fig:2nd_round_demo}  
       \includegraphics[trim={0cm 0cm 0cm 0cm},clip,width=0.75\textwidth]{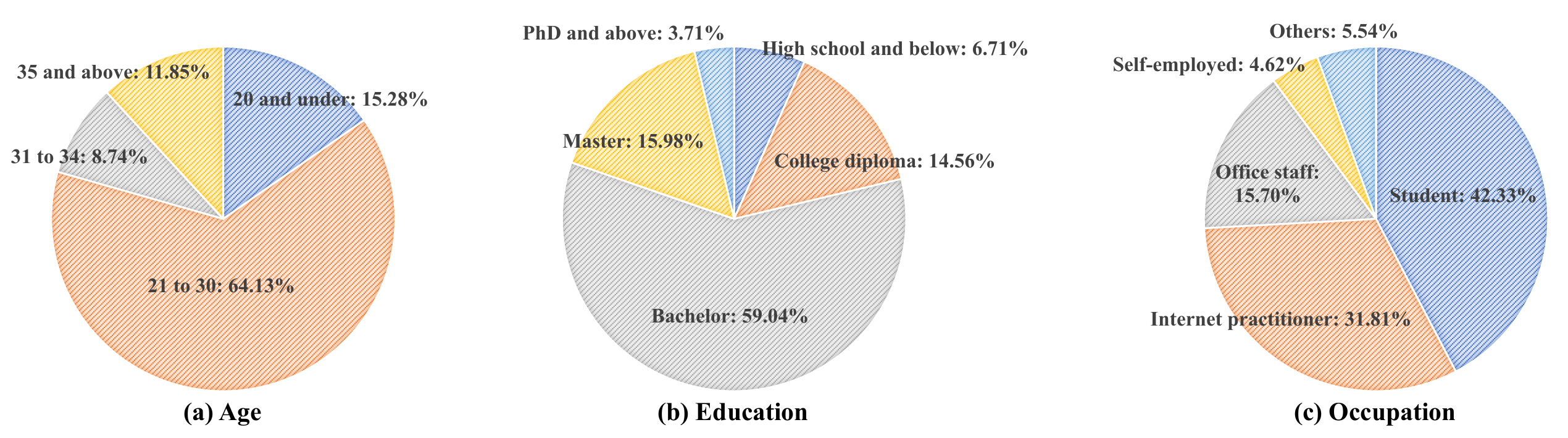}
    }
    \caption{Demographics of respondents from both surveys include age, highest level of education (completed or in progress), and occupation. User profiles remain largely consistent across both surveys.}
    \label{fig:background_stats}
\end{figure*}

\subsubsection{Participants}
The surveys were distributed through CSDN\footnote{Chinese Software Developer Network (CSDN, \url{https://www.csdn.net})}, one of the largest developer communities in China, and promoted on both the CSDN website and its official WeChat account. The first survey ran for 17 days and received 6,644 responses. The second survey ran for five days and received 3,864 responses. We designed two surveys rather than one to balance exploration and validation. The first survey established behavioral signals and tested the validity of our constructs, while the second replicated these measures with an independent sample and extended them to examine longer-term consequences. Together, the two surveys provide both robustness and complementary insights.

To ensure relevance, both surveys included a screening item asking whether respondents were aware of recommender systems in their daily activities. Responses failing this screen were excluded, leaving 5,556 valid responses in the first survey and 2,879 valid responses in the second. Demographic information (age, education, occupation) was collected in each survey. The two samples were consistent in distribution, as shown in Figure \ref{fig:background_stats}. We acknowledge that this recruitment strategy favors technically oriented users, but the large sample sizes and consistency across two independent surveys provide a reliable foundation for our analysis. Demographic information such as age, education, and occupation was also collected, allowing us to assess representativeness and identify potential sampling biases.

\subsubsection{Survey Design}
We discuss how we framed hypotheses that guided our study, the principles of how we designed the survey, and the detailed context for both surveys.

\paragraph{Hypothesis}
\label{ssec:userStudyHypothesis}

We framed four hypotheses to guide our study:

\begin{itemize}
    \item \textbf{Hesitation as a distinct state.} We expect that users often encounter an intermediary state, \textit{hesitation}, before deciding whether they are interested. This state requires additional time and effort compared to immediate acceptance or rejection.
    \item \textbf{Costs of hesitation.} We hypothesize that the effort invested during hesitation influences subsequent judgments of interest or disinterest, shaping the overall experience with recommended items.
    \item \textbf{From hesitation to tolerance.} When hesitation ends in disinterest, we anticipate a secondary state of \textit{tolerance}, marked by frustration over wasted effort. Compared to direct rejection, tolerance represents a more negative and costly user experience.
    \item \textbf{Accumulated impact of tolerance.} Finally, we hypothesize that repeated tolerance erodes trust in the system, reduces engagement, and, over time, drives users toward attrition.
\end{itemize}

These hypotheses highlight the emotional complexity of interacting with recommender systems. By examining states such as hesitation and tolerance, we aim to surface subtle but consequential factors that shape how people experience recommendations and whether they continue to engage with a platform. The hypothesis part in Figure~\ref{fig:conclusion} illustrates this process. Recognizing and addressing these intermediate states opens opportunities to design systems that not only recommend effectively but also respect users' time, reduce frustration, and ultimately foster greater trust and long-term retention.

\paragraph{Survey Design Principles}
\label{ssec:userStudyDesign}

To examine these hypotheses, we designed two large-scale user surveys. Prior work on questionnaire design cautions that asking directly about abstract states, such as hesitation, often produces superficial or unreliable responses ~\cite{krosnick2018questionnaire,schwarz1999self}. Following this guidance, we avoided using the terms \textit{hesitation} or \textit{tolerance} explicitly. Instead, we described realistic scenarios, such as ``assuming interest but realizing otherwise after further exploration,'' to elicit responses that reflect users' lived experiences.

To reduce bias, we conducted two independent surveys at different times. Each was structured into five sections that gradually guided participants from general behaviors toward more reflective judgments. This sequential design encouraged participants to recall experiences step by step, improving both the reliability and validity of the responses.

In the first survey, we focused on capturing how people signal both interest and disinterest in recommendations, as well as the more ambiguous cases where something initially seems appealing but later proves uninteresting. To ground our question design, participants were asked to rate the extent to which they agreed with predefined options, helping us validate the reasonableness of our categories. We also probed whether respondents had experienced hesitation, and invited them to reflect on what triggered it. Finally, we presented scenarios in which initial curiosity gave way to disinterest after hesitation, asking participants to evaluate which of these mirrored their own experiences with recommender systems. 
The second survey replicated the first three sections to cross-check the robustness of our findings, but also introduced two new sections targeting the downstream effects of hesitation and tolerance. These questions examined whether hesitation shaped eventual judgments of interest, and whether repeated tolerance experiences reduced engagement or contributed to attrition.

The two surveys were administered independently at different times, with no intentional overlap in participants. Both included basic demographic items, \textit{i.e.}, age, occupation, and education level, to assess the representativeness of the sample and evaluate the health of the survey traffic. This demographic profiling helped us interpret the validity of the conclusions and ensured that patterns were not artifacts of a narrow subgroup.

\paragraph{First Survey}

The first survey was organized into five sections, each aimed at capturing a different aspect of how people interact with recommender systems (see Appendix~\ref{ssec:firstSurveyQuestions}).

\textit{User profile.} We began by collecting demographic information such as age, education, and occupation. This helped us understand who our respondents were and evaluate the representativeness of the survey platform.

\textit{Context Introduction.} To situate participants, we asked broad questions about their everyday experiences with recommender systems: what they appreciated, what they found frustrating, and how these systems fit into their routines. These reflections established a baseline of perceived value and shortcomings, which informed later sections.

\textit{User Behavior Survey.} We then turned to concrete behaviors in two familiar domains: online shopping and short-video browsing. Participants described what they typically do when they are interested, disinterested, or when something initially seems appealing but later proves not to be. We also asked about incidental actions (\eg curiosity clicks or trending topics) to examine how such behaviors might distort recommendation outcomes.

\textit{Description and Assessment of the ``Hesitation'' State.} Next, we explicitly probed hesitation. Participants were asked whether they had encountered moments of uncertainty and what caused them, such as insufficient information, doubts about quality, or suspicion of advertising. This section encouraged respondents to recall and articulate hesitation as a psychological state in their experience.

\textit{Recognition of Behavioral Combinations.}  Finally, we presented hypothetical scenarios where initial interest gave way to disinterest in shopping and video contexts. By asking users to identify which scenarios matched their experiences, we aimed to validate tolerance behaviors as recognizable and relatable states, rather than abstract constructs.

\begin{table*}
    \centering
     \caption{List of 9 questions ($Q1$ -- $Q9$) and their responses: Users can answer ``yes'' or ``no'' to these questions.  Questions $Q1$ -- $Q3$ are from the first survey, and the remaining questions are from the second survey.}
    \label{tab:surveyQuestions}
    {\small
    \begin{tabular}{cp{5.5in}c}
    \toprule
      \textbf{No.}   & \textbf{Question}  &\textbf{``Yes'' Rate}\\
      \midrule
    Q1  & Have you ever hesitated about information, content, or items recommended to you? (Hesitation refers to being unsure of your interest at the moment and needing additional information or time to make a final decision.)&94\% \\
 Q2& After being recommended a product, if you click to view the item but do not take further actions such as adding it to your cart or making a purchase, does this imply that you are not truly interested in the product?&90\% \\
 Q3& When presented with a piece of content, if you merely skim through it quickly or use the speed-watching feature, result in a viewing percentage much lower than your usual habits, does this suggest that you are not genuinely interested in the content?&88\% \\ 
 Q4& \textbf{Question:} In an online shopping scenario, when you are already well-informed about the basic product information, which of the following actions better reflects a higher level of interest in the product?

\textbf{Option:} You know this is the product you want and place the order directly without much hesitation.&60\% \\  
 Q5&After spending a considerable amount of time viewing recommended content or products and then realizing you are not interested, do you feel frustrated, annoyed, or that your time has been wasted?&59\% \\ 
 Q6&Do these negative emotions become more intense as Q5's increases?&80\%\\ 
 Q7&If the scenario where ``you initially assume you are interested, spend time exploring it, but then realize you are not'' occurs more frequently, would it lower your evaluation of the recommendation feature?&67\% \\ 
 Q8&If the situation where ``you initially assume you are interested, invest time to understand it, only to find out you are not'' becomes more common, would it gradually lead to boredom and prompt you to exit or close the platform?&70\% \\  
 Q9&If you consistently encounter recommendations that fall under the scenario of ``initially assuming interest but realizing after exploration that you are not,'' would you ultimately choose to stop using the platform?&70\% \\ \bottomrule
    \end{tabular}
    }
\end{table*}

\paragraph{Second Survey}

The second survey was designed to complement and validate the first. The initial three sections repeated the same questions as the first survey, allowing us to check consistency across independent samples collected at different times. Beyond replication, two new sections were added to probe more deeply into the \textit{impact} of hesitation and tolerance (see Appendix~\ref{ssec:secondSurveyQuestions}).

\textit{Further Investigation into Hesitation.}
Here, we aim to test the fourth hypothesis in Section~\ref{ssec:userStudyHypothesis}, we asked whether users tend to prefer items they have deliberated over or whether they trust their first impressions and avoid hesitation altogether. This allowed us to test our hypothesis that hesitation is inversely related to satisfaction. We also introduced scenarios of ``assuming interest but realizing otherwise after further exploration,'' preparing participants to reflect on tolerance as a recognizable experience. The scenarios were tailored to both e-commerce and short-video contexts.

\textit{Tolerance and Its Effect on Retention Rates.} This section examined the cumulative consequences of tolerance. Participants were asked whether repeated experiences of wasted time made them less engaged, more likely to disengage, or even inclined to abandon the platform altogether. These questions helped us connect tolerance to longer-term outcomes such as trust, fatigue, and attrition.

We list some of the questions from both surveys in Table~\ref{tab:surveyQuestions}, corresponding to $Q1$ through $Q9$. Respondents can select a \textit{Yes} or \textit{No} answer based on their past experiences with recommender systems. The full list of questions from both surveys is available in Appendix~\ref{sec:surveyDetails}. Due to business reasons, only the survey results (\eg the `Yes' rate as in Table~\ref{tab:surveyQuestions}) for the questions included in the main content of the paper are provided.

\subsubsection{Analysis}
\label{ssec:userStudyAnalysis}

We analyzed survey responses using descriptive statistics for closed-ended questions and thematic analysis~\cite{braun2006using} for open-ended responses. For binary questions, we calculated the proportion of ``yes'' responses to assess prevalence of each experience or behavior. To evaluate consistency across surveys, we compared agreement rates for overlapping questions (Q1-Q3) between the two independent samples. The first three sections were identical across both surveys, and we found highly consistent responses with differences in ``yes'' rates smaller than 2\%. For questions appearing in both surveys, we report results from the first survey due to its larger sample size (5,556 vs. 2,879 valid responses), though both sample sizes provide sufficient statistical power for drawing conclusions.
For open-ended responses describing hesitation triggers and tolerance experiences, we performed thematic analysis by reading through responses, identifying initial patterns, and developing a coding framework. This framework was then applied to categorize responses according to common themes related to our questions about intermediate interaction states.
\subsection{Results and Findings}
\label{ssec:userStudyFindings}
Our findings are organized around four key perspectives: (1) the value proposition of recommender systems, (2) click behavior and user attitudes, (3) presence of hesitation, and (4) the definition and impact of tolerance.
We focus our analysis on two prevalent recommender system contexts: online shopping and content browsing in short-video platforms, as these represent common everyday interactions where users make rapid interest judgments.

\subsubsection{The value proposition of recommender system} 

There are 62\% of respondents favoring recommender systems for optimizing fragmented time use and maximizing the provision of relevant information. Additionally, 48\% express interest in seeing refreshing content that inspires life, while 42\% prioritize saving time in acquiring valuable information. Meanwhile, 56\% of respondents express concerns about privacy, 49\% report dissatisfaction with receiving disinterested, outdated, or repetitive content, and 43\% complain that incidental behaviors have seriously affected recommendation results, leading to irrelevant or incorrect items.

\begin{finding}
High-quality and efficiency are the key values of recommender systems.
\end{finding}

Based on the results above, the primary requirement for recommender systems is to meet users' need for relevant and valuable information during fragmented time slots, even when there is no explicit intention to acquire information.

\subsubsection{Click behavior and user attitudes} 
For online shopping, 68\% of users reported that when they see a product they are interested in, they will click into the detail page for further information. Additionally, 60\% of users said they take further actions on products they are interested in, such as adding them to the shopping cart, bookmarking, or directly purchasing. In the content recommendation channel, 69\% of users said they express interest by reading more content, while 64\% said that further interaction with the platform (such as liking, bookmarking, and sharing with friends) indicates greater interest. Furthermore, 63\% of users expressed long-term interest by following content creators. Conversely, stopping the learning process about an item was the most common way to express disinterest, with 69\% of respondents agreeing. Additionally, using a video speed multiplier or progress bar was identified as a more subtle way of signaling disinterest to the system. Despite these patterns, objective data reveals that 54\% of users who clicked on            ``item details'' did not proceed further, indicating disinterest.

 \begin{finding}
 Clicking is the entry point to learn more about an item, but the action of clicking does not always mean interest.
 \end{finding}
 
In short, while clicking is a common way to acquire more information, it does not necessarily indicate final interest. This highlights the importance of product interface design in accurately interpreting user behavior.

\subsubsection{Presence of hesitation}

The decision-making process requires cognitive engagement from users, and an overly prolonged period of cognitive engagement can be characterized as hesitation. In our survey, only 6\% of users reported never hesitating about the information they were recommended, indicating that hesitation is a very common phenomenon. Then, 64\%  said that hesitation often arises when the current information fails to help them make a judgment or when they have doubts about the perspective and quality of the information. Additionally, 67\% of users reported questioning the authenticity of the information upon seeing words indicating ``advertisement''.

\begin{finding}
    Hesitation is common, and reduces satisfaction with recommender systems.
\end{finding}

\begin{figure}
    \centering
       \includegraphics[width=1\linewidth]{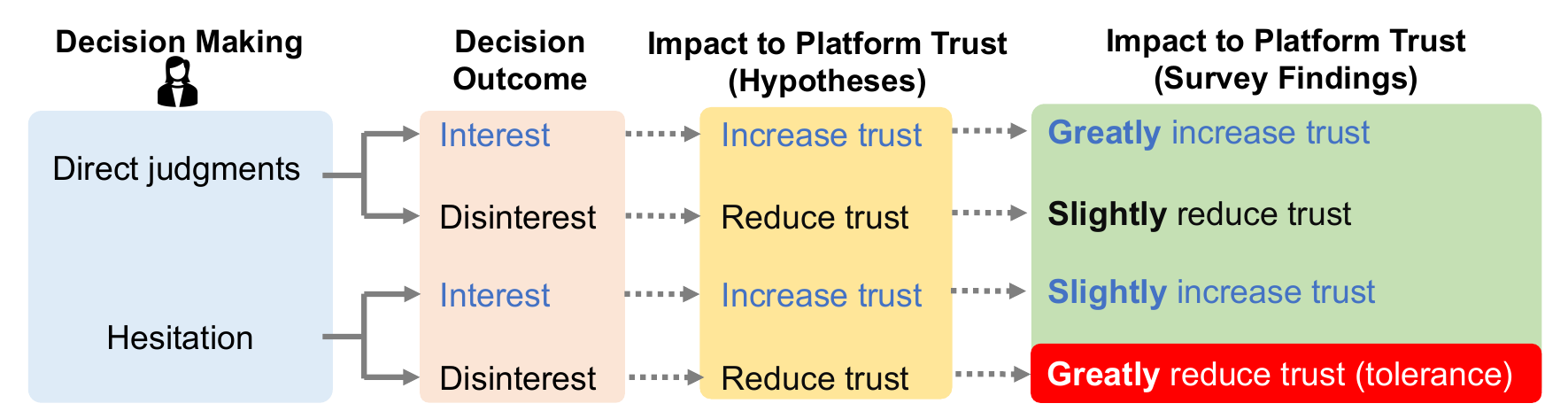}
    \caption{The relationship between users' trust in the platform and their interest in recommended items.}
    \label{fig:conclusion}
\end{figure}

The hesitation process itself increases the user's information acquisition cost, which contradicts the value proposition of recommender systems and affects the user experience. As a result, \textbf{60\%} of respondents indicated a preference for items without a hesitation process, suggesting that hesitation reduces users' interest levels (see Table~\ref{tab:surveyQuestions}). An illustration of our feedback supporting this hypothesis is shown in Figure~\ref{fig:conclusion}.

\subsubsection{The definition and impact of tolerance}
\begin{figure}
    \centering
    \includegraphics[trim={0.8cm 9cm 16.5cm 0.4cm},clip,width=0.8\linewidth]{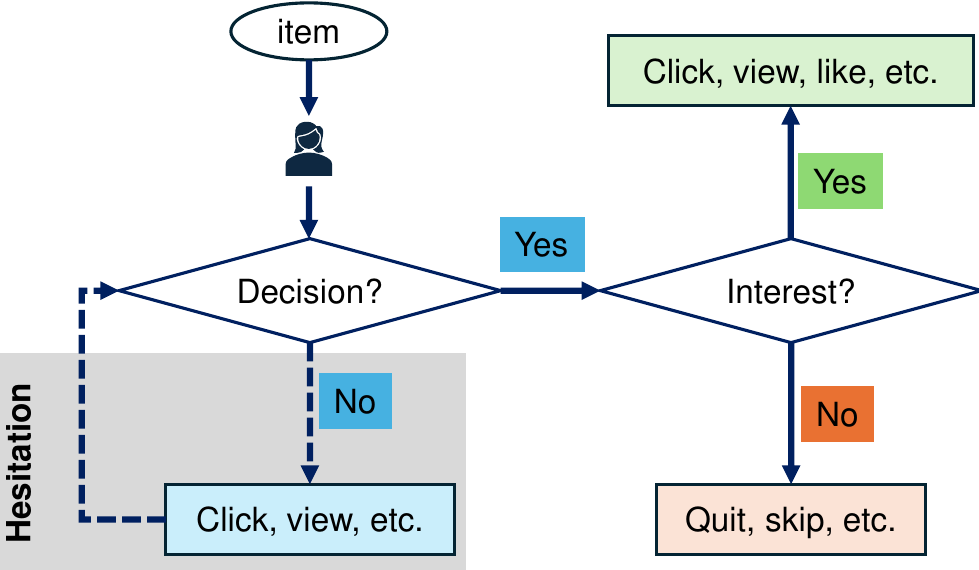}
    \caption{The process of hesitation (and tolerance) on the recommended item/content.}
    \label{fig:t_pic}
\end{figure}

When users experience hesitation followed by a lack of interest, they often display heightened negative emotions due to the wasted time and resources. In our survey, \textbf{59\%} of respondents reported feeling frustrated in such situations, and \textbf{80\%} noted that these negative feelings intensify with repeated experiences (see Table~\ref{tab:surveyQuestions}). Additionally, we observe that a state of tolerance often coexists with indicators of interest but eventually transitions into signals of disinterest, as illustrated in Figure~\ref{fig:t_pic}. More than half of respondents indicated that these behaviors initially sparked their interest, but, upon further exploration, they found the items uninteresting and felt their time had been wasted, a phenomenon we interpret as ``tolerance sessions''.

\textbf{Tolerant behaviors reduce user activity.} From the survey results, \textbf{67\%} of users indicated that as the number of tolerable situations increased, their positive feelings toward the recommender system's functionality decreased, leading them to question its effectiveness. Additionally, \textbf{70\%} of users reported that increased tolerance led to boredom, reduced interaction, and a likelihood of quitting the recommendation function, potentially resulting in platform abandonment.

\textbf{Tolerant behaviors directly affect user retention.} From the survey, \textbf{70\%} of users indicated that if tolerance continues to increase, they would leave the platform and not return. This leads us to an important conclusion of the study: the tolerance phenomenon from the user's perspective directly impacts the platform's retention rate.

\textbf{Typical signals of tolerant behavior.} According to the survey, \textbf{90\%} of users agree that in an online shopping scenario, clicking on a product without taking further actions, such as adding it to the cart, bookmarking, or placing an order, indicates a lack of genuine interest (see Table~\ref{tab:surveyQuestionsE}). Similarly, \textbf{88\%} of users agree that in video browsing, when a video is watched but not for a sufficient amount of time, or is viewed with actions like double-speed playback or skipping, it reflects a lack of interest (see Table~\ref{tab:surveyQuestionsC}). We interpret these behaviors as signs of diminished interest after initial engagement, adding to the user's cognitive load. Consequently, we identify these signals as clear indicators of tolerance under the two surveyed scenarios.

\begin{table}
    \centering
    \caption{User interactions with E-commerce recommender system in different attitudes}
    \label{tab:surveyQuestionsE}
    {\small 
    \begin{tabular}{lp{2.1in}c}
    \toprule
       \textbf{Attitude}& \textbf{User interactions}  &Pct. (\%)\\
      \midrule
     Interest& $\bullet$ Click to view the product details page, examining specifics such as style, model, and reviews.&68\% \\
 & $\bullet$ After reviewing the product details, add the item to cart/favorites or proceed with a purchase.&60\% \\
     \midrule
  Disinterest& $\bullet$ Simply ignore and swipe away the recommendation.&70\% \\  & $\bullet$  Click the ``Not Interested'' option to inform the system to reduce similar recommendations.&65\% \\
  \midrule
 Tolerance& $\bullet$  Click to view the item but do not take further actions such as adding it to cart or making a purchase.&90\% \\
 \bottomrule
    \end{tabular}
}
   
\end{table}

\begin{table}
    \centering
    \caption{User interactions with content-aware recommender system in different attitudes}
    \label{tab:surveyQuestionsC}
    {\small
    \begin{tabular}{lp{2.1in}c}
    \toprule
       \textbf{Attitude}& \textbf{User interactions}  &Pct. (\%)\\
      \midrule
     Interest & $\bullet$ Click to view details and fully explore the recommended content.&69\% \\
 & $\bullet$ Engage with the platform by liking, commenting, sharing with friends, or saving/bookmarking
the content.&64\% \\
 & $\bullet$  Feel inclined to watch more and proactively follow or subscribe to the content creator.&63\% \\
     \midrule
  Disinterest& $\bullet$ Disregard it and quickly swipes it away. &69\% \\  & $\bullet$ Indicate disinterest to the system, requesting fewer similar recommendations.&63\% \\
 & $\bullet$ Immediately switch to other channels or functions, such as using the search feature or checking trending topics.&60\% \\
  \midrule
 Tolerance& $\bullet$ Merely skim through it quickly or use the speed-watching feature, resulting in a viewing percentage much lower than usual habits.&88\% \\
 \bottomrule
    \end{tabular}
   }
\end{table}

\section{Analyzing Tolerance Behavior with Offline Datasets}
\label{ssec:dataset_analysis}

While surveys reveal that hesitation and tolerance erode satisfaction and retention intentions, they remain subjective and subject to recall or reporting biases. Users may claim they would disengage, but whether such disengagement occurs in practice requires objective validation, motivating us to complement them with behavioral datasets. To address this limitation, we analyze large-scale behavioral logs from two contrasting domains, e-commerce (Taobao\footnote{\url{https://tianchi.aliyun.com/dataset/42}}) and short-video (Kuaishou\footnote{\url{https://www.kuairand.com}}). These datasets allow us to systematically observe tolerance behaviors, such as clicking without purchase or watching below one's typical ratio, and examine their correlation with reduced engagement. These behavior signals are user-normalized and reflect systematic under-engagement relative to a user's typical behavior; they therefore distinguish tolerance from random noise or low-quality items. Their consistency across users and their strong predictive value for reduced future engagement further support their interpretation as tolerance signals.

Through this behavioral analysis, we move beyond self-reported perceptions and address our second research question: how can this intermediate state be systematically identified and modeled?

\subsection{Datasets}
\textbf{Taobao} is an e-commerce platform offering millions of items for sale across various categories. User interactions with items on the platform typically involve behaviors such as clicking, adding to cart, or adding to favorites before making a purchase. As discussed in Section~\ref{sec:intro}, clicking on an item signals exploration, but it does not necessarily reflect the user's definitive interest in the item. On an e-commerce platform like Taobao, a purchase is a stronger indication of a user's interest in the item. Similarly, adding an item to the cart or favorites also strongly suggests interest, as it indicates an intent to purchase. In this study, we define \textit{clicking without any subsequent actions} (\eg adding to cart, favorites, or purchasing), as negative feedback from the user. We term this behavior as ``\textbf{tolerance behavior}''. We clarify that tolerance is fundamentally different from \textit{not clicking}. When a user chooses not to click, they demonstrate disinterest without investing time. In contrast, tolerance behavior occurs when a user spends time exploring a product in search of potential interest but ultimately feels that their time was wasted, potentially diminishing their overall experience.

\textbf{Kuaishou} is a video-sharing mobile app that allows users to watch short videos on various topics. However, like Taobao, simply clicking on or watching a video doesn't strongly indicate a user's interests. A better measure is the time spent watching; a longer watch time usually shows greater interest. This watch time also depends on users' browsing habits (\eg patient vs impatient users) and video length (\eg a few seconds vs a few minutes). We define ``\textbf{tolerance behavior}'' as the ratio of a user's watch time to the video's duration. If a user watches a video with a browsing ratio below their average, it is considered tolerance behavior.

\begin{table}
 \small
  \begin{center}
    \caption{Time periods selected for analysis and the corresponding users and items in reference week.}
    \label{tab:taobao_kuaishou_stat}
    \begin{tabular}{@{}l|ccrr}
      \toprule
      \textbf{Dataset}  & \textbf{Ref. Week} & \textbf{Investigation Week} & \textbf{\#User} & \textbf{\#Item} \\
      \midrule
      \multirow{2}{*}{Taobao} & $1$  - $7$, Jun & $8$  - $14$, Jun & $143,475$ & $185,315$  \\
      &  $15$  - $21$, Sep & $22$  - $28$, Sep & $143,514$ & $203,057$  \\
      \midrule
      \multirow{2}{*}{Kuaishou} & $9$  - $15$, Apr & $16$  - $22$, Apr & $26,480$ & $5,429,071$  \\
      &  $23$  - $29$, Apr & $30$ Apr - $6$ May & $26,885$ & $6,001,034$  \\

      \bottomrule
    \end{tabular}
  \end{center}
\end{table}

\subsection{Experimental Setting}
By analyzing tolerance behaviors, we can gain valuable insights into their influence on user experience and, consequently, on retention rates and the overall business potential over time. However, due to the lack of exposure data in the Taobao and Kuaishou datasets, it is infeasible to directly observe user retention within these datasets. Therefore, we focus on \textbf{user engagement} rather than retention rates. 

Engagement serves as a useful indicator of whether a user is likely to continue being active on the platform in the future. Specifically, for Taobao, we use the number of items clicked as a measure of engagement and analyze how the number of clicks without subsequent actions impacts user engagement. For Kuaishou, engagement is measured by the number of videos watched, and we investigate how the browsing ratio affects user engagement.

These proxies are user-normalized and reflect systematic under-engagement relative to a user's typical behavior; they therefore distinguish tolerance from random noise or low-quality items. Their consistency across users and their strong predictive value for reduced future engagement further support their interpretation as tolerance signals.

For both datasets, we select two time periods, each spanning two weeks. Each two-week period is further divided into a ``reference week'' and an ``investigation week.'' During the reference week, we observe user engagement (\ie the number of items clicked on Taobao and the number of videos watched on Kuaishou) and tolerance levels (\ie the number of items clicked without subsequent actions on Taobao and the average browsing ratio on Kuaishou). In the investigation week, we assess whether there is a decline in user engagement compared to the reference week. Table~\ref{tab:taobao_kuaishou_stat} presents the selected time periods along with the corresponding user and item statistics.\footnote{It is worth noting that the selection of time periods is random, and the results remain consistent even when different time periods are chosen.}

\begin{figure}[t]
    \centering
    \subfigure[Taobao]
    {
        \label{sfig:taobao_line}    
        \includegraphics[width=0.4\textwidth]{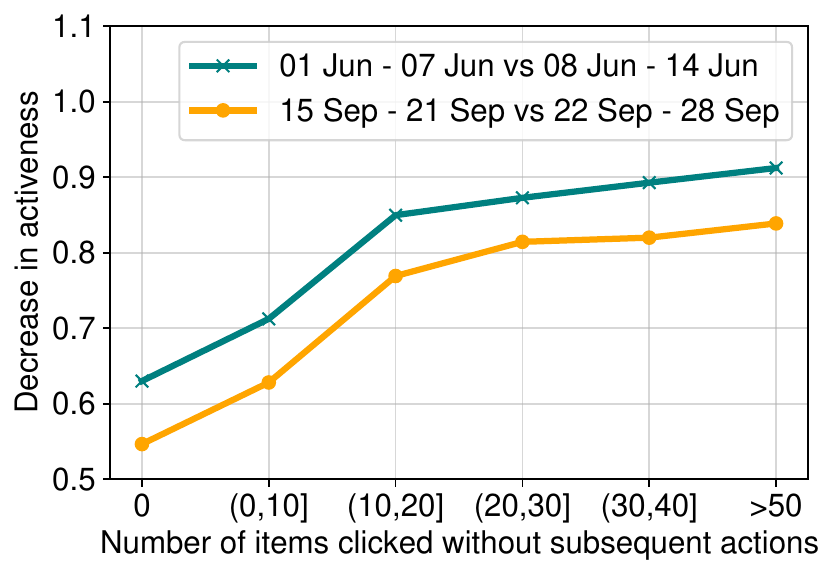}
    }
    \quad
    \subfigure[Kuaishou]
    {
        \label{sfig:kuaishou_line}  
       \includegraphics[width=0.4\textwidth]{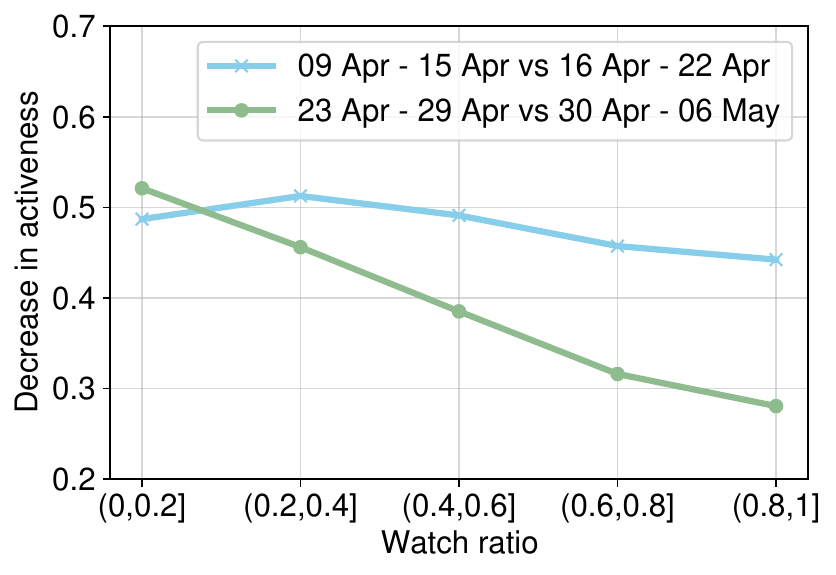}
    }
 
    \caption{Tolerance behaviors in reference week vs proportion of users who show a decrease of engagement in investigation week.}
    \label{fig:tolerance_offline_data}
\end{figure}

\subsection{Tolerance Behavior vs. Engagement}
Figure~\ref{fig:tolerance_offline_data} shows how tolerance behaviors map onto changes in user engagement. On Taobao, people who frequently clicked on products but took no further action were less likely to stay active over time. By contrast, those with relatively few such tolerance events continued to engage more consistently. On Kuaishou, a similar trend emerged: when users watched videos for less than their typical viewing ratio, their overall activity declined.
Across both platforms, the pattern is clear that higher tolerance is strongly linked to reduced engagement. These findings suggest that while tolerance may begin with exploratory actions (\eg clicking or briefly watching), it rarely translates into sustained interest. Instead, tolerance marks a fragile state of attention: users are curious enough to look, but not convinced enough to continue. Over time, this gap erodes satisfaction and increases the likelihood of leaving the platform.

\begin{finding}
    Tolerance is a weak signal of interest, capturing curiosity without commitment, and its accumulation predicts disengagement and eventual attrition.
\end{finding}

\section{Online Experiment: Tolerance Behavior for User Retention}

Based on the findings from our surveys (Section~\ref{ssec:userStudyFindings}) highlight a critical insight: \textit{tolerance behavior toward recommended items undermines user loyalty}. This makes it essential to treat tolerance signals not as noise but as actionable feedback for optimizing recommender systems. 

However, analyses of offline datasets confirm that tolerance behaviors correlate with reduced engagement, but engagement alone only reflects declining interest. Without directly measuring retention, it remains unclear whether modeling tolerance can truly enhance user loyalty. To overcome this limitation, we conduct large-scale online A/B tests on a commercial video platform. By explicitly incorporating tolerance signals into the ranking model and evaluating their impact on next-day retention, we provide causal evidence of their practical value.

Through online experiments, we address our third research question: can modeling tolerance improve retention in practice? By demonstrating causal improvements in retention, a key business metric~\cite{jannach2019measuring}, our study establishes tolerance modeling as an effective and scalable strategy for enhancing recommender performance.

Our initial plan is to conduct both offline and online experiments. Specifically, the offline experiment involves widely used benchmark datasets sourced from e-commerce, video, news, and movie platforms. We aim to design a method that incorporates tolerance signals as a key feature and trains models using historical data over the first $T$ days to predict user exposure on the ${(T+1)}^{th}$ day. In this setting, exposure signifies retention, while a lack of exposure denotes churn.

Unfortunately, existing offline benchmark datasets~\cite{harper2015movielens,gao2022kuairand} do not provide sufficient granularity for predicting whether a user will open the app on day $T+1$. Their user-item interactions are strictly retrospective. This makes it infeasible to observe how retention might change when we differentiate tolerance behaviors from other behaviors driven by interest or disinterest. Given these limitations of offline benchmark datasets, we instead focus on online experiments. Online experiments allow us to explore the intrinsic connection between tolerance signals and user retention more instantly. Meanwhile, online A/B testing in industrial recommender systems enables immediate monitoring of user responses, ensuring that we accurately measure the impact of algorithm modifications on user retention. 

The online A/B experiments were reviewed and approved through both our internal process and the platform's ethics committee. All data used were strictly de-identified and aggregated, ensuring that no personally identifiable information was accessed. In line with platform policies, users could opt out of the influence of experimental recommendation algorithms at any time.

\subsection{Instantiation of Tolerance Behavior}
According to our survey findings, a user's failure to explicitly reject an item does not sufficiently constitute a strong indication of interest or a clear positive signal. Instead, this nuanced response should be regarded as a distinct category of signals that lie outside the realm of clear positivity. Thus, we propose to classify these instances of tolerance behavior as either ``weak positive'' signals or even ``negative'' indicators of user preference. By incorporating this refined categorization, recommender systems gain the capability to interpret user feedback with heightened granularity. It enables recommendation algorithms to better capture the subtle nuances of user interactions and preferences. 

Conventionally, the objective function of a recommendation model is to optimize the binary cross-entropy loss as:
\begin{equation}
    \mathcal{L} = -\sum_{(u, i)\in\mathcal{Y}} y_{ui}\cdot\log \hat{y}_{ui} - \sum_{(u,j)\in\mathcal{Y}^{-}} (1-y_{uj})\cdot\log(1-\hat{y}_{uj})
\label{eq:standard_ce}
\end{equation}
where $\mathcal{Y}$ denotes the positive user-item interaction set and $\mathcal{Y}^{-}$ denotes the negative set. Specifically, $(u,i)$ refers to a user-item pair where user $u$ has interacted with item $i$, such as click on it; $(u,j)$ indicates a user-item pair where user $u$ has not interacted with item $j$, and $y$ is the ground-truth and $\hat{y}$ represents the prediction of recommendation model.

To consider the impact of tolerance signals, we reformulate the conventional objective of recommendation models by introducing the tolerance signals. Specifically, in the video recommendation scenario, we categorize user behavior according to the nature of their engagement. We define the \emph{positive sample} as a \emph{click with follow-up active actions}. For instance, click and watch beyond the user's average viewing percentage.
In contrast, a \emph{tolerance sample} emerges when a user \emph{click without follow-up active actions}, such as click and watch for a duration shorter than the user's average viewing percentage.
Any interaction without a click is labeled as a \emph{negative sample}.
In this case, we further divide the conventional positive signals into positive and tolerable signals. Hence, the positive samples are denoted as $\mathcal{Y}=\mathcal{Y}^{P}\cup\mathcal{Y}^{T}$, where $\mathcal{Y}^{P}$ indicates the clearly positive samples and $\mathcal{Y}^{T}$ indicates the tolerance samples. Through online experiments, we investigate two distinct strategies to handle the tolerance signals.

\textbf{Tolerance samples as negatives}. Here, we treat the tolerance sample as a negative signal and directly merge $\mathcal{Y}^{T}$ and $\mathcal{Y}^{-}$ to form the new set of negative samples. Accordingly, we modify Equation~\ref{eq:standard_ce} to obtain the refined objective as:
\begin{equation}
    \mathcal{L} = -\sum_{(u, i)\in\mathcal{Y}^{P}} y_{ui}\cdot\log \hat{y}_{ui} - \sum_{(u,j)\in\mathcal{Y}^{-}\cup\mathcal{Y}^{T}} (1-y_{uj})\cdot\log(1-\hat{y}_{uj}).
\label{eq:tolerance_ce_neg}
\end{equation}

\textbf{Tolerance samples as weak positives}. A tolerance signal arises directly from a user's clicking behaviors, making it fundamentally distinct from a negative signal (\ie non-click). However, the tolerance signal lacks the robust, follow-up engagement that characterizes a positive signal. As a result, we establish a clear precedence: ``positive signal'' > ``tolerance signal'' > ``negative signal''. To formally incorporate this hierarchy, we employ a discounting mechanism:
\begin{equation}
\begin{aligned}
   \mathcal{L}_T = & -\sum_{(u, i)\in\mathcal{Y}^P} y_{ui}\cdot\log \hat{y}_{ui} - \beta \times \sum_{(u,k)\in\mathcal{Y}^{T}} y_{uk}\cdot\log\hat{y}_{uk} \\ & -\sum_{(u,j)\in\mathcal{Y}^{-}} (1-y_{uj})\cdot\log(1-\hat{y}_{uj}),
\end{aligned}
\label{eq:tolerance_ce}
\end{equation}
where $\beta \in (0,1)$. Here, $\beta$ is a down-weighting coefficient derived directly from how much a user’s viewing duration falls below their own historical completion baseline, after normalization to the interval $(0,1)$. Thus, $\beta$ reflects the degree of under-engagement relative to the user's typical behavior, with larger deviations resulting in smaller $\beta$ values. Note that $\beta$ is not a tunable hyperparameter but a deterministic quantity computed from observed behavior; cases with zero viewing duration are treated directly as negative samples and therefore do not produce $\beta$ values. The key difference between Equations~\ref{eq:standard_ce} and~\ref{eq:tolerance_ce} is that tolerance signals are separated from conventional positives and assigned discounted weights $\beta$, allowing them to be modeled as ``weak positive'' signals. 
These formulations acknowledge that not all clicks are equal: some represent genuine interest, while others capture hesitation or regret.

\subsection{Online Experiments on User Retention}
\label{ssec:exp}

\begin{figure}[t]
    \centering
    \includegraphics[trim={0cm 0cm 0cm 0cm},clip,width=\linewidth]{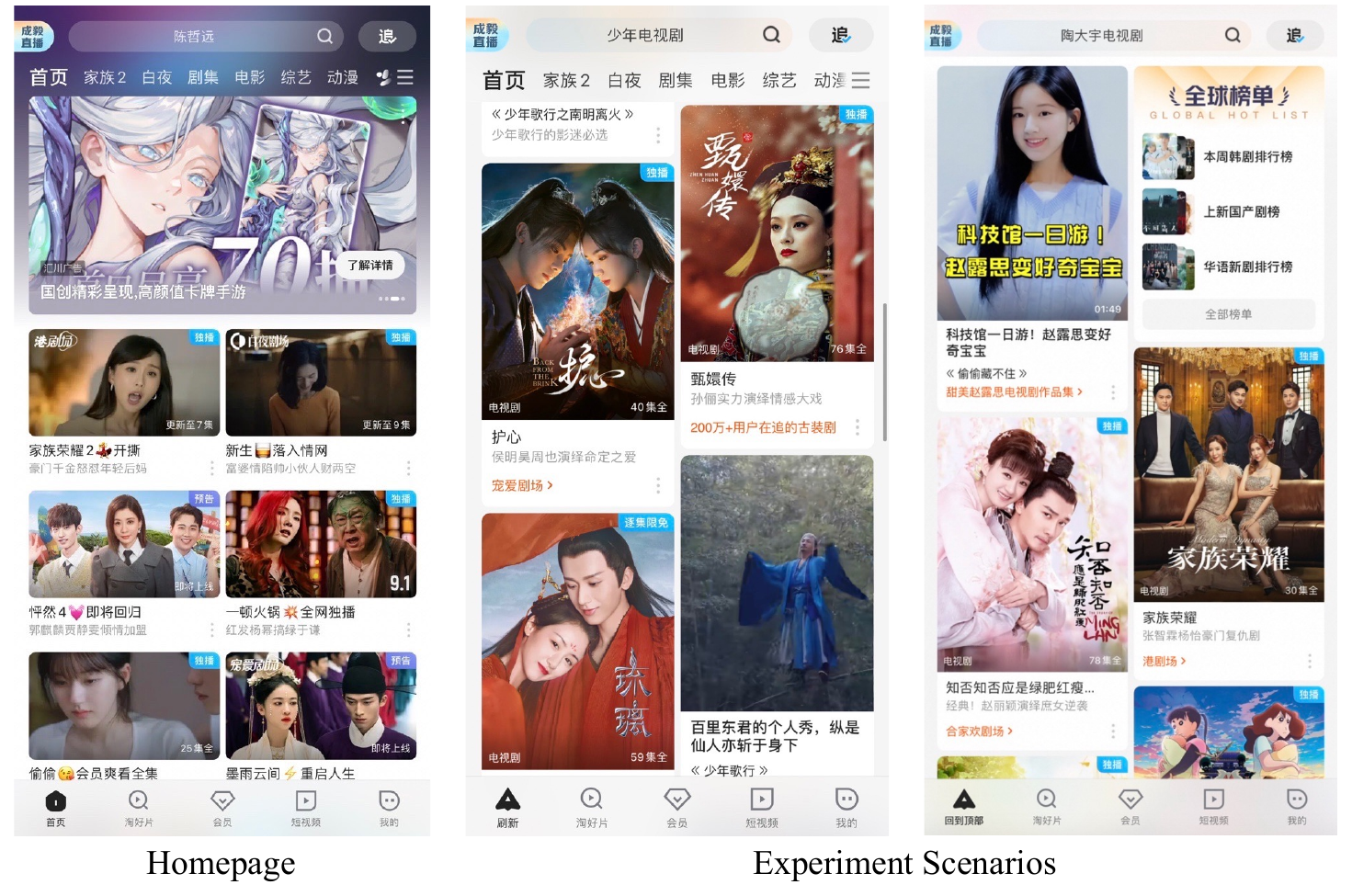}
    \caption{Experiment platform, a Chinese online video and streaming service platform with tens of millions of active users.}
    \label{fig:environment}
\end{figure}

To study the impact of tolerance samples on user retention, we have conducted a series of experiments targeting both optimization objectives in Equations~\ref{eq:tolerance_ce_neg} and~\ref{eq:tolerance_ce} respectively. The results of the experiments aim to reveal the relationship between tolerant behavior and user retention.

\subsubsection{Experiment Setting}
We partner with a Chinese online video and streaming service platform, which has tens of millions of active users, to evaluate our algorithm. The experiments are conducted on the double-column recommendation list on the second screen of the app's homepage, as illustrated in Figure~\ref{fig:environment}. We slightly refine the platform's ranking model (SIM)~\cite{pi2020search} by adjusting the criteria for positive samples based on our findings. Previously, a simple click counted as a positive sample. Now, based on our user study, a positive sample requires a click accompanied by a viewing duration exceeding the user's average viewing percentage. If a user clicks but watches for less time than their average, we label this as a tolerance sample. Situations with no click remain classified as negative samples.

In this study, we conduct two rounds of online A/B testing. Each round included two experimental groups with identical experimental setups but drawn from distinct user cohorts. This approach allows for thorough cross-validation of the results, reducing the risk of bias from a single experiment. In the first round (Groups 1 and 2), we apply Equation~\ref{eq:tolerance_ce_neg}, treating tolerance samples as negative signals that implied user disengagement. In the second round (Groups 3 and 4), we utilize Equation~\ref{eq:tolerance_ce}, interpreting tolerance samples as weakly positive signals that hinted at mild interest or acceptance. Each experimental group comprised 80,000 real users, collectively generating approximately one million page views. The control groups received traffic on a similar scale. 

Our primary evaluation metric is user stickiness, assessed through the Day-2 retention rate change, \ie the percentage of Day 1 active users who remain active on Day 2. Due to potential business changes and competitor influences, conducting long-term experiments is not feasible. Thus, we limit the testing window to seven days. Beyond use retention, we also investigate the impact of tolerance signal isolation on other video recommendation evaluation metrics, such as dwell time.

\subsubsection{Main Results}
Table~\ref{tab:exp_results} summarizes the A/B test results, where $\mathcal{R}$ represents the change in retention rate, \ie the increase in the proportion of users who remain active after being presented with recommended content. The results indicate that adjusting the labels consistently improves the Day-2 retention rate across both independent experimental sessions. This outcome underscores the potential of integrating nuanced behavioral signals into recommender systems to capture fine-grained user preferences and engagement, thereby improving the user retention. It also highlights that more sophisticated behavioral modeling increases user satisfaction and loyalty.
Recent studies have explored the user retention problem in other platforms~\cite{zhu2024interest,10.1145/3604915.3608818,cai2023reinforcing}. For instance, \citet{cai2023reinforcing} modeled user retention optimization as a Markov decision process and developed a retention critic learning, which achieves the retention improvements by around $0.053\%\sim 0.063\%$ in the online A/B test over video recommendation scenario. \citet{10.1145/3604915.3608818} optimized user retention from a contrastive multi-instance learning perspective, which obtains $0.65\%$ next-day user retention improvement over content recommendation scenario.
Despite different online testing platforms, our strategy significantly improves \textit{average 7-day user retention} with \textbf{subtle model modifications or interventions}, reaching $0.67\%$ and $0.36\%$ increase in two separate online tests. These results represent a \textbf{significant improvement for practical business applications given their large user base}, demonstrating our solution's robustness and effectiveness.

\begin{table*}[t]
    \centering
    \caption{Results of four online A/B experiment groups, each tested over a continuous 7-day period.}
    \begin{tabular}{c c c c  c c c c}
    \toprule
    \multicolumn{4}{c}{\textbf{Tolerance as Negative}} & \multicolumn{4}{c}{\textbf{Tolerance as Weak Positive}} \\
    \cmidrule(lr){1-4} \cmidrule(lr){5-8}
    \multicolumn{2}{c}{\textbf{Group 1}} & \multicolumn{2}{c}{\textbf{Group 2}} & \multicolumn{2}{c}{\textbf{Group 3}} & \multicolumn{2}{c}{\textbf{Group 4}} \\
    \cmidrule(lr){1-2} \cmidrule(lr){3-4} \cmidrule(lr){5-6} \cmidrule(lr){7-8} 
    Date & $\mathcal{R}$ & Date & $\mathcal{R}$ & Date & $\mathcal{R}$ & Date & $\mathcal{R}$\\
    \cmidrule(lr){1-2} \cmidrule(lr){3-4} \cmidrule(lr){5-6} \cmidrule(lr){7-8} 
    2024-04-20 & $+0.794\%$ & 2024-04-28 & $+0.670\%$ & 2024-08-21 & $+0.27\%$ & 2024-09-15 & $+1.62\%$ \\
    2024-04-21 & $-0.120\%$ & 2024-04-29 & $-0.694\%$ & 2024-08-22 & $+0.14\%$ & 2024-09-16 & $+0.13\%$ \\
    2024-04-22 & $+1.368\%$ & 2024-04-30 & $+1.695\%$ & 2024-08-23 & $-1.82\%$ & 2024-09-17 & $+1.42\%$ \\
    2024-04-23 & $-0.604\%$ & 2024-05-01 & $+0.576\%$ & 2024-08-24 & $+0.45\%$ & 2024-09-18 & $+0.33\%$ \\
    2024-04-24 & $+0.094\%$ & 2024-05-02 & $-0.132\%$ & 2024-08-25 & $-0.59\%$ & 2024-09-19 & $-0.88\%$ \\
    2024-04-25 & $+0.991\%$ & 2024-05-03 & $+0.045\%$ & 2024-08-26 & $+0.14\%$ & 2024-09-20 & $+2.12\%$ \\
    2024-04-26 & $+1.915\%$ & 2024-05-04 & $+0.417\%$ & 2024-08-27 & $+2.70\%$ & 2024-09-21 & $+0.72\%$ \\
    \cmidrule(lr){1-2} \cmidrule(lr){3-4} \cmidrule(lr){5-6} \cmidrule(lr){7-8} 
    Average & $+0.670\%$ & Average & $+0.360\%$ & Average & $+0.15\%$ & Average & $+0.58\%$ \\
    \bottomrule
    \end{tabular}
    \label{tab:exp_results}
\end{table*}

\begin{table*}[t]
    \centering
    \caption{Changes in the total time users stayed on the platform during the two groups of online A/B experiments.}
    \begin{tabular}{c c c c c c c c}
    \toprule
    \multicolumn{4}{c}{\textbf{Tolerance as Negative}} & \multicolumn{4}{c}{\textbf{Tolerance as Weak Positive}} \\
    \cmidrule(lr){1-4} \cmidrule(lr){5-8}
    \multicolumn{2}{c}{\textbf{Group 1}} & \multicolumn{2}{c}{\textbf{Group 2}} & \multicolumn{2}{c}{\textbf{Group 3}} & \multicolumn{2}{c}{\textbf{Group 4}} \\
    \midrule
    Date & Dwell & Date & Dwell & Date & Dwell & Date & Dwell \\
    \cmidrule(lr){1-2} \cmidrule(lr){3-4} \cmidrule(lr){5-6} \cmidrule(lr){7-8} 
    2024-04-20 & $-1.92\%$ & 2024-04-28 & $-0.79\%$ & 2024-08-21 & $+2.08\%$ & 2024-09-15 & $-2.16\%$ \\
    2024-04-21 & $-1.54\%$ & 2024-04-29 & $+1.19\%$ & 2024-08-22 & $-1.02\%$ & 2024-09-16 & $-2.95\%$ \\
    2024-04-22 & $-3.04\%$ & 2024-04-30 & $-4.63\%$ & 2024-08-23 & $+0.00\%$ & 2024-09-17 & $+0.17\%$ \\
    2024-04-23 & $+1.02\%$ & 2024-05-01 & $+1.33\%$ & 2024-08-24 & $-0.98\%$ & 2024-09-18 & $-3.32\%$ \\
    2024-04-24 & $+1.37\%$ & 2024-05-02 & $-1.48\%$ & 2024-08-25 & $+0.00\%$ & 2024-09-19 & $-3.99\%$ \\
    2024-04-25 & $-0.88\%$ & 2024-05-03 & $-1.81\%$ & 2024-08-26 & $+3.30\%$ & 2024-09-20 & $-0.48\%$ \\
    2024-04-26 & $-3.10\%$ & 2024-05-04 & $-2.77\%$ & 2024-08-27 & $-3.19\%$ & 2024-09-21 & $+0.35\%$ \\
    \cmidrule(lr){1-2} \cmidrule(lr){3-4} \cmidrule(lr){5-6} \cmidrule(lr){7-8} 
    Average & $-1.18\%$ & Average & $-1.28\%$ & Average & $+0.03\%$ & Average & $-1.77\%$ \\
    \bottomrule
    \end{tabular}
    \label{tab:exp_ts_results}
\end{table*}

\subsubsection{Impact on Dwell Time}
The observed gain in retention points to stronger loyalty, yet the accompanying drop in dwell time (see Table~\ref{tab:exp_ts_results})  offers a different perspective on engagement. Traditional recommender systems often equate success with longer sessions, but our findings suggest that duration alone does not guarantee satisfaction. In fact, shorter sessions may reflect the removal of hesitation or indecision --- moments that add length without adding value.
This shift underscores that the \textit{quality} of engagement matters more than its duration. Long sessions only foster loyalty when users feel their time is well spent. Recommender systems should therefore aim to deliver value efficiently, prioritizing content that resonates rather than prolonging use through uncertainty or frustration.
In practice, this means attending to subtle cues such as hesitation and tolerance, and adapting quickly when they appear. By focusing on interactions that minimize wasted effort and maximize relevance, platforms can build deeper trust and long-term engagement, advancing toward more sustainable, user-centered recommendation strategies.

\section{Discussion}
\label{sec:discussion}

Our study identifies \textit{hesitation} and \textit{tolerance} as two intermediate states in user--recommender interactions that are often overlooked when engagement is reduced to clicks and views. Across surveys, behavioral traces, and an online experiment, we find that hesitation is nearly ubiquitous and frequently leads to tolerance, which correlates with frustration, diminished trust, and reduced satisfaction. We now discuss these findings in relation to broader theories of user interaction and intelligent systems, outlining how our conceptualization extends current understandings of how users navigate and interpret recommendations. We then describe how these interaction states challenge long-standing recommender-system assumptions and metrics, followed by limitations and future work.

\subsection{User Interaction States in HCI}
Although we started from a recommender system perspective, this interaction state modeling 
connects to long-standing theories of human action, decision-making, and interpretation.
For example, Norman's framework of gulfs of execution and evaluation \cite{Whitenton2018TwoUXGulfs} helps explain hesitation and tolerance. Hesitation reflects a widened gulf of execution, where users deliberate because it is unclear how or whether to act on a recommendation. Tolerance reflects a gulf of evaluation, where users persist with unwanted content but struggle to assess or act upon their dissatisfaction. Both states reveal frictions that are not captured by binary accept/reject models. 
While related concepts appear in prior work, this study is the first to explicitly model hesitation and tolerance in the context of recommender systems—an increasingly important environment where such frictions often go unexamined.

Our findings also speak to questions of agency and autonomy raised in Value Sensitive Design \cite{friedman1996value, friedman2013value}. Tolerance exemplifies a loss of agency: users continue with unwanted recommendations because exit paths are unclear or costly. Hesitation, meanwhile, illustrates increased cognitive load, as users must invest additional effort to resolve uncertainty. Recognizing these states connects recommender system evaluation to broader theories of effort, engagement, and satisfaction.

Much of our earlier focus emphasized business-oriented metrics such as clicks, dwell time, or retention. Here, we take a step back to foreground the human side of these interactions. By situating hesitation and tolerance within HCI theory, we highlight how user experience, agency, and wellbeing can be compromised when recommendations are designed only for profit. This shift places human value at the center, showing that evaluation must move beyond transactional outcomes to include experiential quality, fairness, and trust --- dimensions that are central to HCI and critical in an era where recommendations mediate so much of everyday life.

Much of our earlier focus emphasized business-oriented metrics such as clicks, dwell time, or retention. Here, we take a step back to foreground the human side of these interactions. By situating hesitation and tolerance within established theories of user experience, agency, and wellbeing, we highlight how these states reveal compromises that arise when recommendations are designed primarily for engagement maximization.

This shift places human value at the center, showing that evaluation must move beyond transactional outcomes to include experiential quality, fairness, and trust --- dimensions that are critical in an era where recommendations shape so much of everyday life.

\subsection{Rethink Clicks and CTR: What Exactly Has Been ``Disrupted''?}
Click-through rate (CTR) has long been treated as the default yardstick for recommender systems. Its popularity stems from convenience, clicks are plentiful, easy to log, and appear to represent interest. For years, both research benchmarks and industry KPIs have revolved around this single number.
Yet our findings complicate this picture. By surfacing the states of \textit{hesitation} and \textit{tolerance}, we show that clicks often capture uncertainty or even dissatisfaction rather than genuine interest. What looks like ``positive engagement'' in the logs may actually be the residue of doubt, wasted time, or frustration, experiences that erode trust and weaken retention.

This calls into question the assumption that \textit{more clicks or longer watch times are always better}. Instead, recommender systems should be optimized for what we term \textit{low-regret satisfaction}: interactions that respect users' time, reduce unnecessary effort, and deliver value that feels meaningful. In other words, engagement metrics must evolve from counting actions to capturing the quality of experience those actions represent.

\subsection{Implication: For Researchers, Designers, Developers, and Users}

\subsubsection{For Researchers: Metrics and Methodologies} 
Our findings suggest that evaluation needs to account for the \textit{process} of decision-making, not just its outcomes. A click should not be treated as a uniform positive signal; rather, it sits within a broader spectrum that includes hesitation, shallow exploration, and tolerance-driven disengagement. By introducing conceptual ``layers'' between pre-click and post-click, we can distinguish strong positives from weak or costly ones. In our video experiments, for example, shallow viewing, defined relative to a person's own viewing baseline, proved to be a reliable indicator of tolerance and produced measurable improvements when modeled explicitly.

Moving forward, we argue that evaluation should adopt metrics that make hidden costs and user burdens visible, rather than relying solely on raw clicks or dwell time. To this end, we propose three new measures:

\begin{itemize}
    \item \textbf{Time-to-Disinterest (TTD).} The elapsed time from exposure to an item until the user explicitly rejects it or exits. TTD captures how long it takes for users to recognize irrelevance, highlighting inefficiencies in the recommendation process.
    \item \textbf{Hesitation Tax.} The additional time cost a user incurs within a session before concluding that they are not interested. This metric surfaces the ``wasted effort'' dimension of hesitation, turning a subjective frustration into a measurable burden.
    \item \textbf{Regret-Adjusted CTR (rCTR).} A click-through rate adjusted by tolerance weights, discounting clicks that lead to shallow engagement or rapid abandonment. rCTR aims to separate curiosity-driven clicks from genuine interest, producing a more faithful indicator of user value.
\end{itemize}
Together, these metrics move evaluation beyond surface-level activity counts, toward measures that acknowledge regret, wasted effort, and erosion of trust.

Equally important is a shift toward long-term objectives. Retention and sustained activity provide a more faithful picture of system impact than immediate clicks or watch time. Our experiments show that even small adjustments, such as reclassifying shallow views, can lead to meaningful gains in next-day retention. This underscores that success should not be defined by \textit{more clicks or longer sessions}, but by cultivating efficient, low-regret interactions that accumulate into durable value over time.

\subsubsection{For Designers: Interaction and Feedback Mechanisms}
Our findings point to several design directions that can help reduce tolerance and protect user trust.

\textit{Make rejection effortless when designing the interactions.} Participants consistently emphasized the frustration of wasted time. Providing quick-access options such as \textit{``Not relevant,'' ``Clickbait,'' or ``Duplicate''} within one tap lowers the cost of signaling dissatisfaction. This prevents tolerance sessions from being misread as positive signals and gives users a sense of control.

\textit{Try to recognize lightweight disengagement.} Everyday actions like fast-forwarding, speeding up playback, skipping ahead, or exiting early should be interpreted as meaningful weak negatives rather than ignored or rolled up into ``watch time.'' Treating these subtle signals as part of the feedback loop aligns system interpretation with actual user intent.

\textit{Let the recommendation session fail fast, adapt quickly.} When tolerance signals appear, the system should pivot, offering alternative content or adjusting its strategy, rather than nudging users to finish what they already find unappealing. Our online experiments confirmed that shortening ``invalid watch time'' does not reduce retention; in fact, it helps maintain satisfaction by respecting users' time. Designing for tolerance means valuing efficiency and agency: letting users say ``no'' easily, treating small disengagements as signals, and responding swiftly when recommendations miss the mark.

\subsubsection{For Developers: Modeling and Engineering}
Our results highlight several opportunities for developers to refine how recommender systems interpret user feedback.

\textit{For modeling, move beyond binary labels.} Instead of treating every click as a positive signal, developers can adopt a richer classification scheme that separates strong positives (deep viewing or follow-up actions), tolerance (shallow or unfulfilled engagement), and clear negatives (no interaction). Modeling these distinctions allows systems to capture subtle but important differences in user intent.

\textit{For user understanding, respect individual baselines.} Whether an interaction counts as tolerance depends heavily on personal habits and content length. A completion rate that looks ``shallow'' for one user may be normal for another. Using relative thresholds --- benchmarked against each individual's own history --- offers a more faithful representation of behavior.

\textit{Digging more data for modeling, like enhancing logging and observability.} To operationalize hesitation and tolerance, platforms need to log more than clicks and dwell time. Events such as fast-forwarding, lingering on a detail page without action, or adjusting playback speed provide valuable signals that can feed into feature stores and model training. Capturing these nuances at scale opens the door to recommendation models that better align with real user experiences.

Engineering for tolerance means treating feedback as multi-layered, personalized, and observable --- bridging raw interaction logs with the complexity of human decision-making.

\subsubsection{For Users: Agency and Well-being}
Finally, our findings highlight the importance of giving users a stronger voice in shaping their recommendations. Many participants expressed frustration when systems misinterpreted exploratory actions as genuine interest, leading to repeated exposure to unwanted content.

\textit{Offer users the right to explanation.} A simple, one-click option such as \textit{``This is not what I expected''} can empower users to clarify their intent. By treating this form of regret-driven feedback as especially meaningful, systems can quickly correct course and avoid reinforcing negative loops. Supporting such mechanisms helps break the chain from tolerance to fatigue to eventual churn. More importantly, it signals respect for users' time and agency, making the recommender system feel less like a black box and more like a partner that listens and adapts.

User-facing controls that acknowledge regret and allow quick clarification build trust, reduce frustration, and prevent tolerance from escalating into abandonment.

\subsection{Limitations for this work}

Our work is not without limitations. First, the scenarios we examined are narrow: our surveys focused on e-commerce and short-video platforms, and our online validation was restricted to one specific placement in a single video app. How tolerance plays out in other domains, such as news, live streaming, or productivity tools, remains uncertain. Second, both our samples and contexts carry potential biases. Survey participants were primarily recruited from developer communities, which may not reflect the diversity of everyday users. Similarly, the online experiments reflect the design choices of one platform and may not generalize elsewhere. Third, our evaluation lens was limited. We focused mainly on Day-2 retention and aggregate watch time, leaving out subjective measures such as satisfaction, perceived regret, or trust, as well as longer-term engagement indicators.

Finally, modeling tolerance touches on ethical questions. Specifically, inferring ``negative emotions'' from behavioral traces risks over-interpretation. Transparency and user agency are essential if such signals are to be used responsibly, ensuring that such modeling empowers rather than exploits users.

\subsection{Future Directions}
There are several promising avenues for extending this work. Larger-scale surveys across more diverse populations and platforms could validate whether our findings generalize beyond the current sample. Beyond breadth, deeper studies are needed to unpack the psychological underpinnings of hesitation, uncertainty, curiosity, and exploratory intent, and to design interventions that are both precise and respectful.

Longer-term experiments, such as rolling A/B tests over weeks or months, would shed light on how tolerance accumulates and affects sustained retention. Cross-domain studies, spanning news feeds, long-form video, live streaming, and multi-stage e-commerce, would test the boundaries of where tolerance matters most.

On the modeling side, future work could integrate tolerance into reinforcement learning or multi-objective optimization, not just as another feature but as a way of shaping rewards around \textit{low-regret engagement}. Comparative studies against purchase-driven ranking or dwell-time-based methods would help clarify its added value.

Another promising direction is to investigate UI-level signals and cues that help users disambiguate interest more efficiently, thereby reducing the likelihood that hesitation escalates into tolerance. Integrating such user interface and experience considerations with model-level tolerance handling could offer a more holistic approach to reducing user regret.

Finally, personalization remains a critical challenge. Adaptive thresholds that reflect individual habits, combined with qualitative approaches such as interviews and field studies, would ground the notion of a ``hesitation tax'' in lived experience. Building shared datasets or benchmarks with tolerance annotations could further accelerate progress, inviting the research community to move beyond clicks and toward metrics that respect time, effort, and trust.

\section{Conclusion}
\label{sec:conclude}
Our study reveals that hesitation is a common part of decision-making in recommender systems. While hesitation signals uncertainty, it often leads users to invest additional effort exploring items --- only to end in disappointment. We define this progression as \textit{tolerance behavior}: interactions that begin with curiosity and engagement but turn into frustration and withdrawal. Tolerance, therefore, offers a powerful signal that current engagement metrics overlook.

By explicitly modeling tolerance in a large-scale video platform, we demonstrated that reframing shallow or regretful interactions as distinct feedback improved next-day retention with minimal engineering cost. This shows that even small adjustments to how we treat user signals can meaningfully shift outcomes, offering both practical scalability and theoretical insight.

More broadly, our findings underscore the need to move beyond clicks and watch time as proxies for ``success.'' Designing recommender systems that minimize wasted effort and respect user time fosters deeper trust, more sustainable engagement, and ultimately more value for both users and platforms.

It is important to acknowledge that our work is subject to certain limitations. We focused on two domains, e-commerce and short video, and validated our modeling approach in only the latter. Whether tolerance manifests similarly in other contexts, such as news, live streaming, or productivity tools, remains an open question. Future work should also explore longer-term outcomes and richer qualitative measures to capture the lived experience of hesitation and tolerance.

\bibliographystyle{ACM-Reference-Format}
\bibliography{RS24Hesitency}

\clearpage

\appendix

\section{Survey Questions}
\label{sec:surveyDetails}

\subsection{The First User survey}
\label{ssec:firstSurveyQuestions}

\subsubsection{Part 1: User Profile Survey }
\begin{itemize}
    \item[\textit{Q1.1}:]  What is your gender? 
    \item[] $\bullet$ Male \quad $\bullet$ Female \quad $\bullet$ Prefer not to disclose
    
    \item[\textit{Q1.2}:] Which of the following best describes your age group?
    \item[] $\bullet$ 20 years old or younger \quad $\bullet$ 21-30 years old   \quad $\bullet$ 31-34  years old\quad $\bullet$ 35 years old or older

    \item[\textit{Q1.3}:] What is the highest level of education you have completed?
    \item[] $\bullet$ High school or below \quad $\bullet$ College diploma \quad $\bullet$ Bachelor's degree \quad $\bullet$ Master's degree \quad  $\bullet$ PhD or higher

     \item[\textit{Q1.4}:] What is your current occupation?
     \item[] $\bullet$ Student \quad $\bullet$ Internet practitioner \quad $\bullet$ Office staff \quad$ \bullet$ Self-employed \quad  $\bullet$ Others
\end{itemize}

\subsubsection{Part 2: Contextual Introduction }

\begin{itemize}

\item[\textit{Q1.5}:] Do you frequently utilize recommended content and related modules (\eg product recommendations, short video recommendations, news suggestions) while using websites and mobile applications?
\item[] $\bullet$ Yes \quad $\bullet$ No 

\item[\textit{Q1.6}:] Is acquiring information through intelligent recommendation channels your primary method of obtaining information when using websites and mobile applications?
\item[] $\bullet$ Yes \quad $\bullet$ No 

\item[\textit{Q1.7}:]  Is purchasing goods through recommendation channels your primary approach to online shopping?
\item[] $\bullet$ Yes \quad $\bullet$ No

 \item[\textit{Q1.8}:] What do you perceive as the most significant value of the recommendation feature?
\item[] $\bullet$ It helps kill time. \newline
     $\bullet$ It saves time that would otherwise be spent searching for information. \newline
     $\bullet$ It provides new and interesting content, offering inspiration or suggestions for daily life.
     
 \item[\textit{Q1.9}:] What frustrates you the most about recommendation channels?
 \item[] $\bullet$ Privacy concerns; excessive collection of personal information by the application. \newline
 $\bullet$ Low quality of recommended content; frequent suggestions of irrelevant, outdated, or repetitive information. \newline
 $\bullet$ Persistent recommendations based on occasional clicks or searches; feeling annoyed without an effective feedback mechanism to avoid such recommendations.
\end{itemize}

\subsubsection{Part 3: User Behavior Survey }
\begin{itemize}

\item[\textit{Q1.10}:] When you are recommended items of interest during online shopping, how do you usually respond?  
\item[] $\bullet$  Click to view the product details page, examining specifics such as style, model, and reviews. \newline 
$\bullet$  After reviewing the product details, add the item to my cart/favorites or proceed with a purchase.

\item[\textit{Q1.11}:] When you are recommended products that are not of interest during online shopping, how do you usually respond? 
\item[] $\bullet$  Simply ignore and swipe away the recommendation. \newline 
$\bullet$  Click the ``Not Interested'' option to inform the system to reduce similar recommendations.\newline 
$\bullet$ Consider the recommendations to be of poor quality and may not to use the app further, opting to close it.

\item[\textit{Q1.12}:] When you encounter interesting recommendations while browsing content, how do you typically respond?
\item[] $\bullet$ Click to view the details and fully explore the recommended content.\newline 
$\bullet$ Engage with the platform by liking, commenting, sharing with friends, or saving/bookmarking the content.\newline
$\bullet$ Feel inclined to watch more and proactively follow or subscribe to the content creator.

\item[\textit{Q1.13}:]  When you encounter uninteresting content while browsing, how do you typically respond?
\item[] $\bullet$ Disregard it and swipe it away.\newline 
$\bullet$ Briefly view or read before confirming it does not match my interest, then stop. \newline 
$\bullet$  Indicate disinterest to the system, requesting fewer similar suggestions. \newline 
$\bullet$ Switch to other channels or functions, such as using the search feature or checking trending topics. \newline
$\bullet$ Close the application.

\item[\textit{Q1.14}:]  If you initially find recommended content interesting but later deem it unappealing upon closer examination, how do you usually respond?
\item[] $\bullet$ Fast-forward to finish viewing the content in a cursory manner. \newline 
$\bullet$ Stop watching without completing the remaining content. \newline $\bullet$ Indicate disinterest in hopes of reducing similar recommendations.

\item[\textit{Q1.15}:]   Do you believe that occasional searches influence subsequent recommendation results?
\item[] $\bullet$ Yes \quad $\bullet$ No 

\item[\textit{Q1.16}:]  If you view content out of curiosity or due to misleading title, then find the content not interesting, do you think this viewing action influences future recommendations?
\item[] $\bullet$ Yes \quad $\bullet$ No 

\item[\textit{Q1.17}:] If you follow trending content due to social influence, despite not having a genuine interest, do you believe this affects future recommendations?
\item[] $\bullet$ Yes \quad $\bullet$ No 
\end{itemize}

\subsubsection{Part 4: Description and Assessment of the ``Hesitation'' State }

\begin{itemize}
    \item[\textit{Q1.18}:]   Have you ever hesitated about information, content, or items recommended to you? (Hesitation refers to being unsure of your interest at the moment and needing additional information or time to make a final decision.) [\textit{Q1 in Table~\ref{tab:surveyQuestions}}]
     \item[] $\bullet$ Yes \quad $\bullet$ No 

    \item[\textit{Q1.19}:]   What are your primary reasons for hesitation?
    \item[]  $\bullet$  Unable to decide based solely on the title or thumbnail image, requiring more details about the product or content. \newline 
    $\bullet$ Having doubts regarding the quality of the product or content (\eg product reviews, authority of the content creator, presentation quality).\newline 
    $\bullet$ The content is labeled as ``Advertisement'' or ``Marketing''.

    \item[\textit{Q1.20}:]   When you feel disinterested in the recommendation results, does it negatively impact your feelings toward this app, to the extent that you may consider switching to other platforms?
    \item[] $\bullet$ Yes \quad $\bullet$ No 
\end{itemize}

\subsubsection{Part 5: Recognition of Behavioral Combination}
\begin{itemize}
    \item[\textit{Q1.21}:] After being recommended a product, if you do not linger beyond the necessary time to view the content, does this indicate that you are not genuinely interested in the product?
    \item[] $\bullet$ Yes \quad $\bullet$ No 

    \item[\textit{Q1.22}:] After being recommended a product, if you click to view the item but do not take further actions such as adding it to your cart or making a purchase, does this imply that you are not truly interested in the product? [\textit{Q2 in Table~\ref{tab:surveyQuestions}}] 
    \item[] $\bullet$ Yes \quad $\bullet$ No

    \item[\textit{Q1.23}:] When presented with a piece of content, if you watch it in its entirety but do not engage with it (\eg through likes, saving, commenting) and immediately move on to the next piece of content, does this suggest that you are not genuinely interested in that content?
    \item[] $\bullet$ Yes \quad $\bullet$ No 

    \item[\textit{Q1.24}:] When presented with a piece of content, if you merely skim through it quickly or use the speed-watching feature, result in a viewing percentage much lower than your usual habits, does this suggest that you are not genuinely interested in the content? [\textit{Q3 in Table~\ref{tab:surveyQuestions}}]
    \item[] $\bullet$ Yes \quad $\bullet$ No 

\end{itemize}

\subsection{The Second Survey}
\label{ssec:secondSurveyQuestions}
Note that \textbf{Parts 1 -3} of the second survey are the same as those in the first survey, and therefore are omitted.

\subsubsection{Part 4: Further Investigation into Hesitation}

\begin{itemize}

     \item[\textit{Q2.18}:] In an online shopping scenario, when you are already well-informed about the basic product information, which of the following actions better reflects a higher level of interest in the product? [\textit{Q4 in Table~\ref{tab:surveyQuestions}}]
    \item[] $\bullet$  Place order  without much hesitation. \newline 
    $\bullet$ Carefully consider whether the product meets my needs, confirming my interest before placing order.

    \item[\textit{Q2.19}:] In the context of online shopping, please select all behaviors that reflect the scenario where initially a recommended product looks interesting to you, but upon further exploration it is not:
    \item[] $\bullet$  View the recommended product but do not take further actions, such as adding it to favorites, adding to cart, or making a purchase. \newline $
    \bullet$ View the recommended product, then click ``Not Interested''. \newline 
    $\bullet$ View the recommended product, then left the recommendation channel for other functions or channels. \newline 
    $\bullet$ View the recommended product, then close the app or webpage afterward. \newline 
    $\bullet$ View the product, add it to favorites or cart, but do not purchase. \newline 
    $\bullet$ Others (please specify).

    \item[\textit{Q2.20}:]  In content browsing scenarios (\eg news, videos), please select all behaviors that  reflect the scenario where a recommended piece of content looks interesting to you, but upon further exploration it is not:
    \item[] $\bullet$ Click on the recommended content but do not finish viewing or reading. \newline  
    $\bullet$ Click on the recommended content but only skim through it. \newline 
    $\bullet$  Click on the recommended content, then click ``Not Interested'' upon finding it not interesting. \newline 
    $\bullet$ Click on the recommended content and, after  finding it not interesting, left the recommendation channel for other channels or functions. \newline
    $\bullet$ Click on the recommended content, upon finding it uninteresting, close the app or webpage. \newline 
    $\bullet$ Others (please specify).
\end{itemize}

\subsubsection{Part 5: Exploration of ``Assuming interested but realize not after further exploration'' and Its Effect on Retention Rate }

\begin{itemize}
     \item[\textit{Q2.21}:]  What is the primary reason for your hesitation?
    \item[] $\bullet$ The title or thumbnail alone is insufficient to make a decision; I need time to check more details about the product or content. \newline  
    $\bullet$  I am unsure whether the product or piece of content is what I need, and need more time to consider it carefully. \newline  
    $\bullet$ I have doubts about the quality of the product or content (\eg product reviews, authority of the content creator, layout quality). \newline 
     $\bullet$ The content is labeled as ``Advertorial'' or ``Marketing''. \newline 
      $\bullet$ Others (please specify).

    \item[\textit{Q2.22}:]  After spending a considerable amount of time viewing recommended content or products and then realizing that they are interesting, do you feel frustrated, annoyed, or that your time has been wasted? [\textit{Q5 in Table~\ref{tab:surveyQuestions}}]
    \item[] $\bullet$ Yes \quad $\bullet$ No 
    
    \item[\textit{Q2.23}:] Do these negative emotions become more intense as Q2.22's increases? [\textit{Q6 in Table~\ref{tab:surveyQuestions}}]
    \item[] $\bullet$ Yes \quad $\bullet$ No

    \item[\textit{Q2.24}:]  If the scenario where ``you initially assume you are interested, spend time exploring it, but then realize you are not'' occurs more frequently, would it lower your evaluation of the recommendation feature? [\textit{Q7 in Table~\ref{tab:surveyQuestions}}]
    \item[] $\bullet$ Yes \quad $\bullet$ No 
    
    \item[\textit{Q2.25}:]  If the situation where ``you initially assume you are interested, invest time to understand it, only to find out you are not'' becomes more common, would it gradually lead to boredom and prompt you to exit or close the platform? [\textit{Q8 in Table~\ref{tab:surveyQuestions}}]
    \item[] $\bullet$ Yes \quad $\bullet$ No 
    
    \item[\textit{Q2.26}:] If you consistently encounter recommendations that fall under the scenario of ``initially assuming interest but realizing after exploration that you are not,'' would you ultimately choose to stop using the platform? [\textit{Q9 in Table~\ref{tab:surveyQuestions}}]
    \item[] $\bullet$ Yes \quad $\bullet$ No

\end{itemize}

\end{document}